\documentclass[11pt]{article}
\usepackage{rotating}
\usepackage{axodraw}
\usepackage{latexsym}
\usepackage{graphicx}
\usepackage{psfrag}
\usepackage{amsmath,amssymb}
\makeatletter
\def\section{\@startsection {section}{1}{\z@}{-3.5ex plus -1ex minus
 -.2ex}{2.3ex plus .2ex}{\large\bf}}
\def\subsection{\@startsection{subsection}{2}{\z@}{-3.25ex plus -1ex
minus -.2ex}{1.5ex plus .2ex}{\normalsize\bf}}
\makeatother
\makeatletter

\@addtoreset{equation}{section}

\makeatother

\def\bea{\begin{eqnarray}} \def\eea{\end{eqnarray}}
\def\be{\begin{equation}} \def\ee{\end{equation}} 
  \def\Z{{\bf Z}}
  
\def\ds{\displaystyle}  
 
  \def\RR{\mathbb R}
\def\mc{\mathcal}

\def\CC{\mathbb C}

\begin{document}

\thispagestyle{empty}

\begin{center}
\hfill SISSA-46/2005/EP \\

\begin{center}

\vspace{1.7cm}

{\LARGE\bf Orbifold resolutions with \\[3mm]
general profile}

\end{center}

\vspace{1.4cm}

{\bf Andrea Wulzer}\\

\vspace{1.2cm}

{\em ISAS-SISSA and INFN, Via Beirut 2-4, I-34013 Trieste, Italy}
\vspace{.3cm}

\end{center}

\vspace{0.8cm}

\centerline{\bf Abstract}
\vspace{2 mm}
\begin{quote}\small

A very general class of resolved versions of the $\CC/\Z_N$, $T^2/\Z_N$ and $S^1/\Z_2$ orbifolds is considered and the free theory of $6D$ chiral fermions studied on it.

As the orbifold limit is taken, localized $4D$ chiral massless fermions are seen to arise at the fixed points. Their number, location and chirality is found to be independent on the detailed profile of the resolving space and to agree with the result of \cite{io}, in which a particular resolution was employed.

As a consistency check of the resolution procedure, the massive equation is numerically studied. In particular, for $S^1/\Z_2$, the "resolved" mass--spectrum and wave functions in the internal space are seen to correctly reproduce the usual orbifold ones, as the orbifold limit is taken.

\end{quote}

\vfill

\newpage

\section{Introduction and Conclusions}

The study of extra--dimensions in a field theory context has received a lot of attention in the last years. This interest, thought originally motivated by string theory ---which somehow "predicts" extra dimensions---, has nowdays gained {\it di per s\'e} motivations, due to the various mechanism which have been proposed to address some of the theoretical issues of the Standard Model (SM), such as the Plank/Weak hierarchy problem \cite{RS12}, or to provide a more elegant and unified framework (such as gauge unification \cite{gauge-uni}) or, at a more phenomenological level, to address the pressing problem of Little Hierarchies \cite{BHN}. The final goal of these works would be to find an extra--dimensional effective field theory, whose effect could be already visible in near future experiments, "completing" the SM in the ultraviolet. A recent interesting proposal is contained in \cite{ACP}.

In this context, compactification on orbifolds \cite{Dixon} is a particularly useful tool. Simple orbifolds, in particular $S^1/\Z_2$ and $T^2/\Z_N$ (see {\it e.g.} \cite{ghu6d}), are frequently employed in model building, for their capability of breaking symmetries and introducing chirality in the fermion spectrum. Both these effects, however, could be also obtained by flux compactification, which consists on considering more complicated non--flat spaces on which background for the gauge (and eventually gravity) field strength is present. The main reason for preferring orbifolds is simplicity; when considering fluxes, indeed, it becomes technically much more involved (see {\it e.g.} \cite{ib}) to deal with the resulting theory, while an orbifold model can be studied on the covering space ---which is in general trivial and flat such as the circle or the torus--- by gauging away the discrete ($\Z_N$) orbifold symmetry. Apart of simplicity, another important reason why orbifolds are so commonly used is the flexibility in their $4D$ field content. At the fixed points, indeed, $4D$ localized fields of any kind are commonly introduced in arbitrary number.

Although their apparent simplicity, orbifolds are singular, the singularities being located at the fixed points, and can be seen as singular limits of smooth "resolving" spaces. Under this point of view, orbifold compactification is just flux compactification, in the limit in which the fluxes approach a singular profile. In a recent paper \cite{io}, a resolved version of the $\CC/\Z_N$, $T^2/\Z_2$ and $S^1/\Z_2$ orbifolds was constructed, and the free theory of a $6D$ chiral spinor was studied on them. It was there shown that each orbifold singularity admits various topologically different resolutions, labelled by an integer monopole charge $q$ one can add at the resolved fixed point. As a consequence of this fact, the index of the Dirac operator on the resolved space, which counts the number of $L$--handed minus the number of $R$--handed zero modes and depends on these monopole charges, as implied by the Atiyah--Singer theorem, can assume different values, so that different fermion zero--modes spectra can arise on the resolved theory. This may appear to be an inconsistency, since in the un--resolved orbifold model, which the resolution should resamble in the suitable limit, the number and chirality of the zero--modes which survive the orbifold projection is fixed. The study of the resolved Dirac equation performed in \cite{io} reveals, however, a very nice physical interpretation of the extra zero--modes. Apart of the few of them which are constant, that exactly correspond to the usual "bulk" orbifold ones, their wave function is found to be peaked around the resolved singularity, and its probability density to become a delta--function in the orbifold limit. A certain, quite wide, number of "brane" fermion distributions are then found to originate naturally, as an effect of the resolution, from a single "bulk" $6D$ field.

In \cite{io}, for simplicity, a particular shape of the resolving space was assumed, such that the Dirac equation could be analytically solved on it. As a resolution of the $\CC/\Z_N$ orbifold, which corresponds to a cone in the $\RR^3$ embedding, a ${\mc C}^1$ space obtained by gluing a truncated cone with a suitable portion of a sphere was considered. The resolution of the compact $T^2/\Z_2$ was obtained by gluing four such spaces to a flat "bulk" region. A "resolved version" of $S^1/\Z_2$ was constructed by attaching two halves--spheres to a cylinder. On these spherical caps, a background for the gauge connection $A=\kappa/2\omega$ was introduced, proportional to the spin connection $\omega$, the coefficient of proportionality $\kappa$ depending on the arbitrary monopole charge $q$. Removing this assumption and generalizing the results of \cite{io} to arbitrary (possibly ${\mc C}^\infty$) deformations of the above described resolutions is one of the results of the present paper. This demonstrates that the phenomenon of fermion localization, and the distributions of localized states one gets, do not depend on the particular resolution performed but, on the contrary, are intimately connected with the nature of the resolved orbifold. However, since various mechanisms for fermion localization have been studied in the literature \footnote{In  \cite{localization}, fermion localization was originally shown to arise on topological defects. See also  \cite{localization-new,localizationFI} for a more recent example in an extra--dimensional context.} , the present analysis does not exclude that other patterns of localized fermions, different from those found here, could be realized. What is peculiar of the present results is that they are obtained by considering the minimal background which is required for resolving the singularity, with fermion fields which are minimally coupled to it. Note that, even though orbifold field theories may appear to need a resolution ---at least at the level of regularization--- when dealing with localized operators of particular types, there is no true physical need for resolving the singularities. The un--resolved orbifold models can be safely considered as effective field theories, the scale of the resolution (if any) being above (or at) its physical cut--off. It is then important to remark that the phenomenon of fermion localization observed here is completely independent on the resolution scale, which can be arbitrarily high. The "allowed" brane fermion configurations listed in this paper can be directly considered in the unresolved orbifold model, as a remnant of the high energy resolution. All other effects related to the finite size of the singularities can be neglected. 

The paper is organized as follows. In Sect.~2, the theory of a fermion field on $\CC/\Z_N$ is considered; the derivation of massless and massive fermion wave functions is quickly reviewed. The general resolution of the above model, compatible with the $O(2)$ isometry group of $\CC/\Z_N$ and with the simplest space--time topology, is constructed in Sect.~3. In Sect.~4, the Dirac equation on the resolved orbifold is studied, and the wave functions compared with the orbifold ones. Independently on the detailed profile of the resolution, the number and chirality of the massless zero--modes can be extracted. The independence of the zero--modes spectrum on the detailed profile of the resolution, even for the compact $T^2/\Z_N$ orbifolds, whose resolution simply consists on resolved $\CC/\Z_N$ cones attached to a flat bulk region, is then demonstrated. In the massive case, the Dirac equation is numerically solved when a particular class of ${\mc C}^\infty$ resolving profiles, labelled by a parameter $\delta$, is employed. The wave functions are shown to reproduce, in the appropriate limit and independently on $\delta$, the orbifold ones. This provides a check that the resolved theory, independently on its detailed profile, really mimic the orbifold model. In Sect.~5, an analogous study is performed for $S^1/\Z_2$, whose general "resolution" is introduced and the massless and massive Dirac equations analyzed. The number and chirality of the zero--modes is derived, the mass--spectrum and the wave functions are shown to reproduce the orbifold ones. In Appendix A, two dimensional spaces with $O(2)$ isometry are discussed. In Appendix B, the resolution of the $T^2/\Z_N$ compact orbifolds is defined, and the spectrum of localized zero--modes is derived. This generalize \cite{io}, in which only the case of $T^2/\Z_2$ was considered.

\section{Fermions on $\CC/\Z_N$}

The $\CC/\Z_N$ orbifolds are obtained from the complex plane $\CC$ with euclidean metric by identifying points connected by a $2\pi/N$ rotation around the origin. The fundamental domain one gets is the $2\pi/N$ plane angle in between the $z=t$ and $z=\tau\,t$ ($t$ real an positive, $\tau=e^{2\pi i/N}$) lines, which are identified. The plane angle can be deformed isometrically in $\RR^3$ until the two extremal lines  coincide, the resulting surface being a cone of angle $\alpha$, with $\sin{\alpha}=1/N$. Aim of this section is to solve the Dirac equation for a $2D$ spinor field
\be
\psi=\left(\begin{array}{c}\psi_R\\ \psi_L \end{array}\right)\,,
\label{sde}
\ee
endowed with an $U(1)$ symmetry, on $\CC/\Z_N$. It is defined as a field on $\CC$ which remains invariant under a $2\pi/N$ rotation around the origin, modulo a suitable phase transformation
\be
\psi(\tau z)={\mc P}\psi(z)\,,\;\;\;\;\;\;\;\;{\mc P}={\ds e^{\pi i\left(1-\frac1N\right){\sigma_3}}
e^{\frac{\pi i}{N}p} }\,,
\label{bcN}
\ee
where $p$ is any integer running from $-N+1$ to $N-1$ at steps of $2$.
The $U(1)$ phase in Eq.~(\ref{bcN}) is chosen to make ${\mc P}^N=1$ and an extra $-1=e^{\pi i\sigma_3}$ has been included in the Lorentz part of ${\mc P}$ for future convenience. The Dirac equation on $\CC$, in the complex coordinates $z$ and $\bar z$, reads \footnote{Here and in the following, Dirac equations on $2D$ euclidean spaces ${\mc E}_2$ will be considered. These equations, however, also represent the eigenmode equations for the wave functions in the internal space of $6D$ chiral fermion fields on ${\mc M}_4\times {\mc E}_2$. The $L$ and $R$ subscripts in Eq.~(\ref{sde}) then also refer to the $4D$ chirality of the corresponding $4D$ fields.}
\be
\left\{
\begin{array}{l}
2\partial_z\psi_L=-im\psi_R\\
2\partial_{\bar z}\psi_R=-im\psi_L
\end{array}
\right.\,.
\label{eqC}
\ee
and diplays its invariance under the $O(2)$ isometry group of rotations and parity. Note that, in general, parity is broken by the orbifold condition in  Eq.~(\ref{bcN}). The parity symmetry is only preserved when $p=0$. By defining polar coordinates $z=\tau_c e^{i\theta}$ ($\tau_c\in(0,\infty)$, $\theta$ is an angle), Eq.~(\ref{eqC}) can be rewritten as
\be
\left\{
\begin{array}{l}
\left(i\partial_{\tau_c}+\frac1{\tau_c}\partial_\theta\right)\psi_L=me^{i\theta}\psi_R\\
\left(i\partial_{\tau_c}-\frac1{\tau_c}\partial_\theta\right)\psi_R=me^{-i\theta}e\psi_L
\end{array}
\right.\,,
\label{eqplane}
\ee
and its $SO(2)$ symmetry corresponding to Lorentz rotations acts as
\be
\theta\;\rightarrow\;\theta-\beta\,,\;\;\;\;\;\psi\;\rightarrow\;\ds{e^{\frac i2\beta\sigma_3}}\psi\,.
\label{sy}
\ee
The operator $i\partial_\theta-{\sigma_3}/2$, which generates the above symmetry, can be diagonalized with real eigenvalues on the space of solutions to the Dirac equation. Therefore, one can look at solutions of the form
\be
\psi_R=f_R(\tau_c)e^{i\mu\theta}\,,\;\;\;\;\;\psi_L=f_L(\tau_c)e^{i(\mu+1)\theta}\,,
\label{anm}
\ee
with $\mu$ integer. For $m\neq0$, then, the ansatz (\ref{anm}) will be used to parametrize the $\theta-$dependence of the fields. For $m=0$, the symmetry of the Dirac equation is enhanced, since it becomes invariant under $\psi\rightarrow e^{i\beta\sigma_3}\psi$ transformations also. One can simultaneously diagonalize the $\sigma_3$ and $i\partial_\theta$ operators and the ansatz in this case is more general:
\be
\psi_R=f_R(\tau_c)e^{i\mu_R\theta}\,,\;\;\;\;\;\psi_L=f_L(\tau_c)e^{i\mu_L\theta}\,,
\label{an0}
\ee
with $\mu_{L,R}$ arbitrary integers.

Eq.~(\ref{eqplane}) for $m=0$, when the ansatz (\ref{an0}) is inserted, becomes a set of two trivial decoupled first order differential equations whose solutions are given by 
\be
f_R={\tau_c}^{\mu_R}\,,\;\;\;\;\;f_L={\tau_c}^{-\mu_L}\,.
\label{orsolm0}
\ee
Since wave functions of the form (\ref{an0}) must vanish (or at most be finite, if $\mu_{L,R}=0$) at $\tau_c=0$, to be single-valued at that point, one must impose $\mu_R\ge0$ and $\mu_L\le0$. Moreover, if one restricts the search of solutions to those which are finite for $\tau_c\rightarrow\infty$, the cases $\mu_{L,R}=0$, corresponding to constant zero-modes, are the only allowed ones. To get the $\CC/\Z_N$ wave functions, the orbifold condition in Eq.~(\ref{bcN}) must be imposed. Vanishing $\mu_{R,L}$ is allowed just for $p=\mp (N-1)$, in which case the $R(L)$--handed part of the orbifold twist becomes the identity. In the massive case, the ansatz (\ref{anm}) must be inserted in Eq.~(\ref{eqplane}), then leading to the first order system of coupled differential equations
\be
\left\{
\begin{array}{l}
\left(\partial_{\tau_c}+\frac{\mu+1}{\tau_c}\right)(if_L)=mf_R\\
\left(\partial_{\tau_c}-\frac\mu{\tau_c}\right)f_R=-m(if_L)
\end{array}
\right.\,,
\label{eqmor}
\ee
whose general solution is a linear combination of first and second kind Bessell functions
\be
\begin{array}{l}
f_R(\tau_c)=A\,J_\mu(m\tau_c)+B\,Y_\mu(m\tau_c)\,,\\
if_L(\tau_c)=A\,J_{\mu+1}(m\tau_c)+B\,Y_{\mu+1}(m\tau_c)\,.
\end{array}
\label{solorm}
\ee
Bessell functions of both kinds asymptotically vanish for large positive value of the argument and the ones of first kind ($J_n(z)$) are regular for $z=0$ while those of second kind ($Y_n(z)$) are not. The only regular (and finite at infinity) solutions to the massive Dirac equation on $\CC$ are then expressed by Eq.~(\ref{solorm}) with $B=0$. Among those, the orbifold condition (\ref{bcN}) selects only the ones for which 
\be
\mu=\frac{p+N-1}2+Nl\,,
\label{almu}
\ee
for some integer value of $l$.

\section{General resolutions of $\CC/\Z_N$}

The $\CC/Z_N$ orbifold is a cone of angle $\alpha$ with $\sin{\alpha}=1/N$.
The line element of this space can be written as
\be
ds^2=d\tau_{c}^{2}+\ds{\sin^2\alpha}\,\tau_{c}^2d\phi^2\,,
\label{cle}
\ee
in the polar coordinates $(\tau_c,\phi)$, which are related to the complex plane one used in the previous section by $z=\tau_c\ds{e^{i\frac\phi{N}}}$. Of course, a space with the metric (\ref{cle}), if $\sin{\alpha}\neq 1$, is singular at $\tau_c=0$, and then needs a resolution. The cone possesses an $O(2)$ isometry group, whose $\RR^3$ embedding consists on rotations around its axis ($\phi\rightarrow\phi+\lambda$) and reflections orthogonal to 
any plane which contains it ($\phi\rightarrow-\phi+\lambda$). It is very reasonable to 
assume the resolving space ${\mc R}$ to possess the same isometry group of 
the space it has to resolve. Moreover, only spaces which can be entirely described with a single set of coordinates, and are then topologically trivial, will be considered in the following. These assumptions, though very reasonable, exclude of course more complicate resolutions, in which ${\mc R}$ breaks the $O(2)$ isometry, which is however restored in the orbifold limit, or it has handles, and then non-trivial topology, but they still leave a very general class of spaces. Topologically trivial smooth $2D$ spaces with $O(2)$ isometry group can be parametrized (see Appendix~A) in terms of a single function $\rho(\tau)$, $\tau\in(0,\infty)$, which completely defines the metric, once expressed in the polar coordinates $(\tau,\,\phi)$. In these coordinates
\be
ds^2=d\tau^2+\rho^2(\tau)d\phi^2\,,
\label{metric}
\ee
where $\rho(\tau)$ has to satisfy certain constraints (see Eq.~(\ref{con})) ensuring that the space so defined is really regular at $\tau=0$, where polar coordinates are ill-defined. If ${\mc R}$ has to reproduce the cone far enough from the singularity, at, say $\tau\ge\eta$, its metric (\ref{metric}) must reduce to the one in Eq.~(\ref{cle}) up a coordinate change. Therefore, for $\tau\ge\eta$, one requires
\be
\rho(\tau)\equiv\sin{\alpha}(\tau+\tau_{c}^0)\,,
\label{rs}
\ee
in such a way that, identifying the coordinate $\tau_c$ in Eq.~(\ref{cle}) with $\tau+\tau_{c}^0=\tau_c$, ${\mc R}$ reproduces, when $\tau>\eta$, a portion ($\tau_c\in(\eta+\tau_{c}^0,\infty)$) of a cone. Since $\rho\sim\tau$ for $\tau\sim0$ (see Eq.~(\ref{con})), the role of the resolving space is to interpolate a plane at $\tau\sim0$ with a cone at $\tau\ge\eta$. For ${\mc R}$ to be tractable, {\it i.e.} the tangent plane to be well defined, the first derivative of $\rho$ must be at least continuous.  A resolution with $\rho\in{\mc C}^1$ at $\eta$ is called a "${\mc C}^1$ resolution", and an example was provided in \cite{io} consisting in a spherical cap attached to a truncated cone. A class of simple ${\mc C}^\infty$ profiles that will be useful in the following, parametrized by a positive real number $\delta$, is defined, for $\tau\le\eta$, as
\be
\rho(\tau)=\int_{0}^\tau d\tau'\left[
\sin{\alpha}+(1-\sin{\alpha})\ds{e^{\delta^2 \left(
\frac1{\eta^2}+\frac1{{\tau'}^2-\eta^2}\right)}}
\right]
\,.
\label{cinpr}
\ee

Aim of this section is to define a resolved version of the theory, described in Sect.~2, of a free fermion on $\CC/\Z_N$. Consider then a fermion field $\psi$ on ${\mc R}$. Apparently, one essential feature for this model to be related with the orbifold one would be that, in the space region where ${\mc R}$ coincides with the cone ($\tau>\eta$), the $\psi$ field on ${\mc R}$ and the one on $\CC/\Z_N$ should share the same boundary conditions, provided by Eq.~(\ref{bcN}), under $\phi\rightarrow\phi+2\pi$ ($z\rightarrow\tau z$). Note that, however, the boundary conditions alone cannot have any intrinsic meaning. Indeed, they can be changed and eventually made trivial by field redefinitions, consiting on non-periodic $U(1)$ gauge or Local-Lorentz (LL) transformations which, at the same time, also affect the gauge field and the spin connection, turning on non-trivial (flat) backgrounds $A$ and $\omega$ for them. What really matters, on the contrary, is the {\em holonomy} of the field on circuits surrounding the singularity. To the holonomy, which is gauge invariant, both the boundary conditions and the background of the gauge and spin connections contribute. In the case of trivial boundary conditions, it is expressed by the appropriate Wilson loop. In the case of the orbifold model, in which the connections vanish, one has
\be
W=W_{LL}W_{gauge}={\mc P}^{-1}=\ds{e^{-\pi i\left(1-\frac1N\right)\sigma_3}e^{-\frac{\pi i}N p}}\,.
\label{holo}
\ee
Since the resolving space ${\mc R}$ is entirely described with a single coordinate set, the fields on it have to be described by single-valued functions. The fermion will then be periodic as $\phi\rightarrow\phi+2\pi$ and backgrounds for $A$ and $\omega$ must be present. Note that the presence of a non-trivial $LL$ connection $\omega$ is automatic from Eq.~(\ref{metric}) and its form can be shown to be consistent with  Eq.~(\ref{holo}). The detailed form of the gauge background $A$ is, on the contrary, arbitrary. The connections $A$ and $\omega$ are globally defined ($O(2)$-invariant, consistently with what was assumed to be the isometry group) vector fields on ${\mc R}$, with negative intrinsic parity under the $\Z_2$ ($\phi\rightarrow-\phi$) action. Note that the coupling of fermions to the $A$ background breaks the parity symmetry (which interchanges $L$-- and $R$--handed fields), while the one to $\omega$ does not. This is correct since, as observed in Sect.1, it is the presence of the $U(1)$ gauge part in the orbifold twist matrix ${\mc P}$, which in turn is related to the gauge part of the holonomy and then requires $A$ to have a v.e.v., that makes the fermions violate parity. Any globally well defined covariant (pseudo-)vector $\Omega$ (see Appendix~A) can be parametrized in terms of a single function of $\tau$, its $\phi$ component $\Omega_\phi$; $\Omega_\tau$ being equal to zero. As for $\rho(\tau)$, there are some conditions on $\Omega_\phi$ and its derivatives at $\tau=0$ (see Eq.~(\ref{00})). These conditions are automatically satisfied by the spin connection, once the appropriate local frame has been chosen. Starting from the metric in Eq.~(\ref{metric}), it is straightforward to derive the $2$-bein forms $\widehat\theta^\alpha$
\be
\widehat\theta^\alpha=\left(\ds{e^{-i\phi\sigma_2}}\right)^{\alpha}_{\;\beta}\widehat\theta^{\beta}_0=
\left(\begin{array}{l}\cos{\phi}d\tau-\sin{\phi}\rho(\tau)d\phi\\
\sin{\phi}d\tau+\cos{\phi}\rho(\tau)d\phi
\end{array}\right)\,,
\label{2b}
\ee
and, by imposing the torsion-free condition, find the associated spin connection 
\be
\omega=(1-\dot\rho(\tau))d\phi\,.
\label{om}
\ee
Note that $\omega$, as expected, becomes a pure gauge ($R=d\omega=0$) for $\tau\ge\eta$, thank to the condition (\ref{rs}). The $LL$ Wilson line $W_{LL}$ on fermions is immediately computed. It is $W_{LL}=\ds{e^{-\frac i2\oint \sigma_3\omega}}=\ds{e^{-\pi i \sigma_3\left(1-\sin{\alpha}\right)}}$, which precisely matches what is needed for reproducing the $LL$ holonomy of the orbifold in Eq.~(\ref{holo}). Since a vanishing field-strength background is present in the bulk, the gauge connection $A$ must become a pure gauge for $\tau\ge\eta$
\be
A(\tau)\equiv \frac\kappa2 (1-\sin{\alpha})d\phi\,,\;\;\;\;\;\kappa=\frac{p+2N q}{N-1}\,,
\label{kap}
\ee
where $\kappa$ has been chosen so that $W_{gauge}=\ds{e^{-i\oint A}}$ satisfies Eq.~(\ref{holo}). An arbitrary integer "monopole" charge $q$, to whose presence the holonomy is insensitive, has been included in Eq.~(\ref{kap}). When a concrete example of resolution will be needed, the gauge connection $A$ will be taken to be proportional to $\omega$: $A=\kappa/2\omega$.

Summarizing, the general resolution of the orbifold model of Sect.1 is parametrized by two arbitrary functions, $\rho(\tau)$ and $A_{\phi}(\tau)$, constrained to satisfy Eq.s~(\ref{con}),(\ref{rs}),(\ref{00}) and (\ref{kap}). The resulting space, with its gauge field background, is equivalent by construction to the $\CC/\Z_N$ orbifold for $\tau\ge\eta$. The precise relation between fermion fields on ${\mc R}$ and the orbifold ones is given by
\be
\psi_c(\tau_c,\theta)=\ds{e^{\frac{i}2(N-1)(\kappa+\sigma_3)\theta}}\psi(\tau_c-\tau_{c}^0,N\theta)\,.
\label{or-m-fer}
\ee
Clearly, Eq.~(\ref{or-m-fer}) makes a non--trivial twist under $\theta\rightarrow\theta+2\pi/N$ appear in $\psi_c$ which precisely matches Eq.~(\ref{bcN}). Note that all the results of the present section are based on necessity arguments. Checks will be provided in the following that the resolved theory really mimic, at energy much below the resolution scale $1/\eta$, the orbifold one.

\section{Fermions on general resolutions}

Given the $2$-beins in Eq.~(\ref{2b}), the Dirac equation is easily written
\be
\left\{\begin{array}{l}
\partial_\tau\psi_L-\frac i{\rho(\tau)}\left[\partial_\phi+i(A_\phi-\frac12\omega_\phi)\right]\psi_L=-ime^{i\phi}\psi_R\\
\partial_\tau\psi_R+\frac i{\rho(\tau)}\left[\partial_\phi+i(A_\phi+\frac12\omega_\phi)\right]\psi_R=-ime^{-i\phi}\psi_L
\end{array}\right.\,.
\label{de}
\ee
Clearly, it possesses an $SO(2)$ invariance, which acts as in Eq.~(\ref{sy}) with $\theta$ replaced by $\phi$. The ansatz for $m\neq0$ will then be
\be
\psi_R=e^{in\phi}f_R(\tau)\,,\;\;\;\;\;\psi_L=e^{i(n+1)\phi}f_L(\tau)\,,
\label{ans0}
\ee
with $n$ integer. For $m=0$, the ansatz is changed to
\be
\psi_{L,R}=f_{L,R}(\tau)e^{i n_{L,R}\phi}\,,
\label{ansm}
\ee
with $n_{L,R}$, of course, integers.

\subsection{Zero modes}

With the ansatz (\ref{ans0}), the Dirac equation (\ref{de}) for $m=0$ becomes
\be
\left\{\begin{array}{l}
\partial_\tau \log{f_R}=\frac1{\rho(\tau)}\left[n_R+A_\phi+\frac12\omega_\phi\right]\\
\partial_\tau \log{f_L}=-\frac1{\rho(\tau)}\left[n_L+A_\phi-\frac12\omega_\phi\right]
\end{array}
\right.\,.
\label{eq0}
\ee
It is impossible, of course, to integrate Eq.~(\ref{eq0}) without specifying a particular shape of $\rho(\tau)$ and $A_\phi(\tau)$. The behavior of the solutions for $\tau\sim0$ and $\tau\ge\eta$, however, is universal, since universal is the form of $\rho$ and $A_\phi$ in these regions. For $\tau\sim0$, $\rho(\tau)\sim\tau$, $A_\phi(\tau)\sim0$ and Eq.~(\ref{eq0}) approximates the one on the $\CC$ plane already encountered in Sect.~2. The solutions at $\tau\sim0$ behave then as
$$
f_{R}(\tau)\sim\tau^{n_R}\,,\;\;\;\;\;f_{L}(\tau)\sim\tau^{-n_L}\,,
$$
which means the only well-behaved solutions to be those with $n_R\ge0$, $n_L\le0$. In the $\tau\ge\eta$ region, $\rho\equiv\sin{\alpha}(\tau+\tau_{c}^0)$, $A_\phi\equiv\kappa/2(1-\sin{\alpha})$. Eq.~(\ref{eq0}) then becomes, in the coordinate $\tau_c=\tau+\tau_{c}^0$, the Dirac equation on the cone, whose solutions are, up to a constant
\be
f_{R}={\tau_c}^{\lambda_R}\,,\;\;\;\;\;f_{L}={\tau_c}^{-\lambda_L}\,,
\label{wfr}
\ee
with the integers $\lambda_{L,R}$ defined as
\be
\begin{array}{l}
\lambda_R=\lambda_R(n_R)=\frac1{\sin{\alpha}}(n_R+\frac{\kappa+1}2(1-\sin{\alpha}))=\frac{p+N-1}2+N(n_R+q)   \,,\\
\lambda_L=\lambda_L(n_L)=\frac1{\sin{\alpha}}(n_L+\frac{\kappa-1}2(1-\sin{\alpha}))=\frac{p-(N-1)}2+N(n_L+q)  \,.
\end{array}
\label{lamLR}
\ee
By comparing Eq.~(\ref{or-m-fer}) with Eqs.~(\ref{ans0}), (\ref{an0}) and Eq.~(\ref{orsolm0}) with Eq.~(\ref{wfr}), one immediately sees that each fermion state on ${\mc R}$ with angular momentum $n_{L,R}$ reproduces the orbifold one (see Sect.~2) with angular momentum $\mu_{L,R}=\lambda_{L,R}(n_{L,R})$. In the case of the orbifold, however, the only allowed states are those with $\mu_R\ge0$ and $\mu_L\le0$ while any value for $\lambda_{L,R}$ is allowed now for which $n_R\ge0$ and $n_L\le0$ in Eq.~(\ref{lamLR}). In Sect.~2, when restricting to wave functions which are finite or zero for $\tau\rightarrow\infty$, the only surviving solutions were found to be the constant ones ($\mu_L=0$ or $\mu_R=0$), in the cases $p=N-1$ or $p=-(N-1)$, respectively. On the resolved orbifold, on the contrary, there are in general several values of $n_R\ge0$ or $n_L\le0$ for which $\lambda_R\le0$ or $\lambda_L\ge0$, meaning that extra solutions are found whose number crucially depends on the value of the monopole charge $q$ introduced in Eq.~(\ref{kap}). Not for any value $q$, however, the constant $\CC/\Z_N$ wave functions can be found among the ones on ${\mc R}$. It is indeed clear from Eq.~(\ref{lamLR}) that, for $p=-(N-1)$, a vanishing $\lambda_R$ can be realized with $n_R\ge0$ only if $q\le0$, while $\lambda_L=0$ requires $p=(N-1)$ and $q\ge0$. This means that, when $p$ reaches its maximal and minimal values, a particular sign of $q$ must be chosen for the resolved theory really mimic the orbifold one.

The orbifold interpretation of the extra massless states is clear; they correspond to "brane" ($4D$--chiral) fermion fields localized at the fixed point of the $\CC/\Z_N$ orbifold. Note that, in any case but when $|p|$ is maximal and $q$ has the "wrong" sign, all these states are indeed \emph{bound states}, since they can be normalized to $\int\psi^\dagger\psi=1$, and their shape is \emph{localized} around the singularity, so that $\psi^\dagger\psi$ reproduces a delta function in the orbifold limit. Fermion localization arising from orbifold resolution was discussed in \cite{io}, where Eq.~(\ref{lamLR}) was derived with a particular choice of the profile of the resolving space, and used for studying the zero--modes on the resolved $T^2/\Z_2$ orbifold, obtained by gluing four resolved $\CC/\Z_2$ cones together. Eq.~(\ref{lamLR}) is the only information on the resolution which is needed for working out the zero--modes. Having shown now that Eq.~(\ref{lamLR}) is universal, {\it i.e.} independent on the profile of the resolved space, therefore means that the results of \cite{io}, and in particular the allowed patterns of localized fermions there computed, are intimately connected with the nature of the resolved orbifold, instead of being an accident depending on the particular resolution performed.

In Appendix B, Eq.~(\ref{lamLR}) will be used to work out the zero--modes on the $T^2/\Z_N$ orbifolds,  generalizing the results of \cite{io}.

\subsection{The massive Dirac Equation}

With the ansatz (\ref{ansm}), Eq.~(\ref{de}) becomes 
\be
\left\{\begin{array}{l}
\partial_\tau(if_L)+\frac 1{\rho(\tau)}\left[n+1+(A_\phi-\frac12\omega_\phi)\right](if_L)
=mf_R\\
\partial_\tau f_R-\frac 1{\rho(\tau)}\left[n+(A_\phi+\frac12\omega_\phi)\right]f_R=-m(if_L)
\end{array}\right.\,.
\label{eqm}
\ee
For $\tau\sim0$ it reduces, as expected, to the one on the $\CC$ plane in Eq.~(\ref{eqmor}) with $\mu$ replaced by $n$ and $\tau_c$ by $\tau$. The general solution to this equation is given in Eq.~(\ref{solorm}). Among that solutions, however, only the ones with $B=0$ are finite at $\tau=0$. Any regular solution of Eq.~(\ref{eqm}), for $\tau\ll\eta$, assumes then the form
\be
f_R= J_n(m\tau)\,,\;\;\;\;\;if_L= J_{n+1}(m\tau)\,,
\label{zerobe}
\ee
where an irrelevant common multiplicative factor has been set to one. At $\tau\ge\eta$, again, Eq.~(\ref{eqm}) becomes of the form (\ref{eqmor}) and its general solution is given by 
\be
\begin{array}{l}
f_R(\tau)=A(m,\eta)\,J_\lambda(m\tau_{c})+B(m,\eta)\,Y_\lambda(m\tau_{c})\,,\\
if_L(\tau)=A(m,\eta)\,J_{\lambda+1}(m\tau_{c})+B(m,\eta)\,Y_{\lambda+1}(m\tau_{c})\,,
\end{array}
\label{etabe}
\ee
with $\tau_c=\tau+\tau_{c}^0$ and having defined the integer
\be
\lambda=\lambda(n)=\frac1{\sin{\alpha}}\left(n+\frac{k+1}2(1-\sin{\alpha})\right)=\frac{p+N-1}2+N(n+q)\,.
\label{lamm}
\ee
Using Eq.~(\ref{or-m-fer}) to compare Eq.~(\ref{anm}) with Eq.~(\ref{ansm}) one easily recognizes that any massive fermion on ${\mc R}$ with a given angular momentum $n$ should correspond to the massive state on $\CC/\Z_N$ with orbifold angular momentum $\mu=\lambda(n)$. By comparing Eq.~(\ref{lamm}) with Eq.~(\ref{almu}) one sees that the allowed values of $\lambda$ are precisely those which the orbifold condition allows for $\mu$.
If ${\mc R}$ has to reproduce the orbifold, the massive wave function (\ref{etabe}) has to reduce, up to an irrelevant multiplicative factor, to the orbifold one, with $B=0$, for $m\ll1/\eta$. What is then to be computed is the ratio ${\mc r}(m,\eta)\equiv B(m,\eta)/A(m,\eta)$, which should be seen to go to zero as $m\eta\rightarrow0$. Unfortunately, no analytical profile--independent informations on ${\mc r}(m,\eta)$ can be extracted, even in the small $m\eta$ limit, since $A$ and $B$ in Eq.~(\ref{etabe}) crucially depend on how the solution evolves, due to Eq.~(\ref{eqm}), from $\tau\sim0$, where consistency requires the "initial condition" (\ref{zerobe}), to $\tau=\eta$. The best can be done, therefore, is to solve numerically Eq.~(\ref{eqm}) from $0$ to $\eta$ when a particular profile has been chosen and compute ${\mc r}$, for $m\eta$ smaller and smaller, by matching Eq.~(\ref{etabe}) with the exact (numerical) solution. \footnote{The author thanks M.Neri for her help in implementing the required software. Various routines developed in Numerical Recipes \cite{num.res} for differential equation solving, numerical integration and computation of Bessell functions have been employed. See Footnote 5 for some additional detail on the numerical procedure followed.} The class of profiles in Eq.~(\ref{cinpr}), and $A=\kappa/2\omega$, has been used. The behavior of ${\mc r}(m,\eta)$ for $m\eta\ll 1$ is seen to depend on the parameter $\delta$ which characterizes the profile, the limit being however profile--independent.

As an example, consider the orbifold $\CC/\Z_3$ with maximal gauge twist $p=2$, and its resolution with monopole charge $q=1$. As discussed in the previous section, since $q$ has the same sign as $p=N-1$, two $L$--handed massless states are present; one is constant and corresponds to the zero--mode of the orbifold bulk fermion, the other is "localized" near the origin and should be interpreted as a brane massless fermion.
Focus on the particular massive state with $n=-2$ (whose orbifold angular momentum is $\lambda=-1$), mass $m=0.1/\eta$ and resolution parameter $\delta=\eta$. Eq.~(\ref{eqm}) can be numerically solved from $0$ to $\eta$; the coefficients $A$ and $B$ of Eq.~(\ref{etabe}) are extracted by imposing continuity at $\tau=\eta$. In the case at hand, one finds ${\mc r}\sim2\,10^{-2}$, quite small if compared with $m\eta=0.1$. By decreasing the mass one finds, for $m\eta=0.01$, ${\mc r}\sim2\,10^{-4}$ and for $m\eta=10^{-3}$, ${\mc r}\sim2\,10^{-6}$. The ratio ${\mc r}$, then, is proportional to $(m\eta)^2$ in this case. It is possible to verify that sensible changes of the resolution parameter $\delta$ weakly modify the situation. For $\delta=10^{-2}\eta$, $m\eta=0.1$, for instance, ${\mc r}\sim5\,10^{-2}$. When considering an angular momentum $n\neq-2$ the situation is also better. For the same orbifold theory and the same $q$ as before, with $\delta=\eta$ and $m\eta=0.1$, one gets ${\mc r}\sim7\,10^{-5}$ for the state $n=-1$ and ${\mc r}\sim1\,10^{-17}$ for the one with $n=0$. Changing the monopole charge, if it remains positive, the situation does not change significantly. For $N=3$, $p=2$ as before, but $q=2$, one gets ${\mc r}\sim2\,10^{-2}$ for $m\eta=0.1$ and $n=-3$, and smaller results for other angular momenta.

The situation drastically changes when $q$ becomes negative, $p$ being positive. Consider again $p=2$ and $N=3$, but $q=-1$. In this case, as discussed in the previous section, only one $R$--handed localized massless state is present, and there is no track of the $L$--handed bulk state of the original orbifold. This means that the orbifold theory, in this particular case, is not correctly reproduced by the resolution. A signal of this fact, indeed, can be also retrieved from studying the massive states. Take $n=0$, $\delta=\eta$ and $m\eta=0.1$; one finds ${\mc r}\sim8\,10^{-1}$ which is significantly larger than the values found up to now. Moreover, for $m\eta=10^{-2},\,10^{-3},\,10^{-4}$ one finds ${\mc r}\sim4\,10^{-1},\,2\,10^{-1},\,1\,10^{-1}$ respectively. The convergence of the bulk solution to the orbifold one is, then, only logarithmic in this case. This fact is common to any orbifold $\CC/\Z_N$ with $p=N-1$, $q<0$, for the state with $\lambda=-1$. In the case $p=-(N-1)$, $q>0$, it is the state with $\lambda=0$ which displays logarithmic convergence. Finally, when $|p|$ is not maximal, both signs of $q$ are allowed, since ${\mc r}$ always diplays power-like convergence to zero.

\section{The "resolved" $S^1/\Z_2$ orbifold}

The one--dimensional orbifold $S^1/\Z_2$, much more than the two--dimensional ones considered up to now, is of great phenomenological importance, since many models have been formulated in which it has been used to describe the internal dimension. Geometrically, $S^1/\Z_2$ is simply a line segment, and the two points of the circle which are fixed under the $\Z_2$ action are simply \emph{boundaries} and then, differently from the two--dimensional cases considered up to now, are not singular points at all. At the purely geometrical level, therefore, there is no reason for trying to replace $S^1/\Z_2$ with some  "smoothed version" of it. Thought doubts could be aroused on the full consistency of field theories on a segment, one may argue that these technicalities do not signal any physical need of resolving the orbifold effective theory, which may be considered as it is. The approach followed here, however, is weakly related to the above considerations. The resolution may just be seen as a mechanism for the localization of fermions, that will be shown to take place in this case as in the two--dimensional ones. The final result will consist on a list of "special" brane fermion distributions, to be eventually studied in the unresolved orbifold theory, which can be obtained from considering $S^1/\Z_2$ as a limit of a boundaryless $2D$ space. Clearly, introducing one more dimension in the resolving theory slightly restricts the class of $5D$ models one can treat. Only those can be considered whose field content can be interpreted as arising from a $6D$ model. $5D$ Dirac fields will be reproduced by starting from chiral $6D$  spinors.

\subsection{The resolving space}

The space ${\mc C}$ which resolves the $S^1/\Z_2$ segment, as proposed in \cite{io}, is taken to be a $2D$ compact ''cigar-like'' surface which resembles, in a certain region, a finite portion of a cylinder of radius $r$. The cylinder becomes, in the orbifold limit in which $r$ shrinks to zero, a line segment which reproduces the bulk of $S^1/\Z_2$. The rest of the space consists on two disconnected regions and each of them will shrink to a point in the orbifold limit. They provide the resolved description of the orbifold fixed points. The topology of ${\mc C}$ is assumed to be as simple as possible, {\it i.e} the one of the sphere. Namely, each fixed point will be described by a single chart, the overlapping of the two being provided by the cylinder. Moreover, the $O(2)$ isometry group of the cylinder is taken to be the isometry of the whole space. The results of Appendix A can then be applied in each coordinate system, and the metric parametrized as
\be
(ds^2)_i=d{\tau_i}^2+{\rho_i}^2(\tau_i)d{\phi_i}^2\,,
\label{fpi}
\ee
where $\phi_{1,2}$ are angles and $\tau_{1,2}$ both run in the $[0,\,L-\eta]$ interval, being $L$ the total "length" of ${\mc C}$, {\it i.e.} the distance between its two "poles" $\tau_1=0$ and $\tau_2=0$. The two coordinate systems are related by $\tau_2=L-\tau_1$ and $\phi_2=-\phi_1$ in the overlapping region $\tau_i\in[\eta,L-\eta]$ which parametrizes the cylinder. The resolved fixed points are described by the two disconnected regions $\tau_{1,2}\in [0,\eta]$, and the orbifold limit consists on taking  $\eta\rightarrow0$. Since each $(\tau_i,\,\phi_i)$ coordinate system is ill--defined at $\tau_i=0$, $\rho_i(\tau_i)$ and its derivatives must satisfy certain conditions at $\tau_i=0$, summarized in Eq.~(\ref{con}), ensuring that no physical singularity is present at that point. Moreover, since Eq.~(\ref{fpi}) must reduce to the flat cylinder metric: $\rho_{1,2}(\tau_i)\equiv\rho_{1,2}(\eta)\equiv r$ for $\tau_i\ge\eta$. As a particular class of ${\mc C}^\infty$ profiles for $\rho_i$, consider
\be
\rho_i(\tau_i)=\int_{0}^{\tau_i} d\tau\left[\ds{e^{\delta^2 \left(
\frac1{\eta^2}+\frac1{\tau^2-\eta^2}\right)}}
\right]
\,.
\label{cinprs1}
\ee

Be $\psi$ a spinor field on ${\mc C}$. It will be described, in each coordinate system, by single--valued functions, so that its holonomy on circles wrapping around the cylinder is entirely given by the appropriate Wilson loop. The Local-Lorentz part of the holonomy is easily computed: $W_{LL}=e^{-\pi i\sigma_3}=-1$. The total holonomy needs to be the identity, if some light state has to survive when the orbifold limit is taken. A non trivial gauge background $A^i$ ($i=1,2$ labels the two coordinate systems) must then be included to generate a gauge holonomy $W_{gauge}=-1$. This background, as usual, is assumed to be $O(2)$ invariant and then (see Appendix A) it has the form $A^i=A^{i}_{\phi_i}(\tau_i)d\phi_i$, and is  subjected to the conditions (\ref{00}) at $\tau_i=0$. Moreover, $A^i$ must to reduce to a pure gauge for $\tau_i\ge\eta$,
\be
A^i=\frac{\tilde\kappa_i}2d\phi_i\,,
\label{gai}
\ee
where $\tilde\kappa_i$ needs to be an odd integer if requiring $W_{gauge}=e^{-i\oint A}=-1$. Having chosen $A^{1,2}$ in Eq.~(\ref{gai}), the gauge $+$ $LL$ transformations relating, on the cylinder, the representations of the fermion field in the two coordinate systems is fixed to be
\be
\psi^2(L-\tau_1,-\phi_1)=\ds{e^{i\left(\sigma_3+\frac{\tilde\kappa_1+\tilde\kappa_2}2\right)\phi_1}
e^{-i\frac\pi2\sigma_3}}\psi^1(\tau_1,\phi_1)\,.
\label{cooch}
\ee
The Dirac equation on ${\mc C}$ is given, in each coordinate system, by Eq.~(\ref{de}) and, as in Eq.~(\ref{ans0}), its $SO(2)$ invariance can be used to parametrize the $\phi_i$--dependence of massive states as
\be
\psi_{R}^i=f_{R}^i(\tau_i)\ds{e^{in_i\phi_i}}\,,\;\;\;\;\;
\psi_{L}^i=f_{L}^i(\tau_i)\ds{e^{i(n_i+1)\phi_i}}\,.
\label{ancim}
\ee
In the massless case, as in Eq.~(\ref{ansm}), the ansatz is
\be
\psi_{L,R}^i=f_{R,L}^i(\tau_i)\ds{e^{in_{L,R}^i\phi_i}}\,,
\label{anci0}
\ee
with $n_{L,R}^i$, and $n_i$, integers.

\subsection{Zero-modes}

With the ansatz (\ref{anci0}), the massless Dirac equation reads
\be
\left\{\begin{array}{l}
\partial_{\tau_i} \log{f_{R}^i}=\frac1{\rho_i(\tau_i)}\left[n_{R}^i+A_{\phi}^i+\frac12\omega_\phi\right]\\
\partial_{\tau_i} \log{f_{L}^i}=-\frac1{\rho_i(\tau_i)}\left[n_{L}^i+A_{\phi}^i-\frac12\omega_\phi\right]
\end{array}
\right.\,.
\label{de0cig}
\ee
At $\tau_i\sim0$ it reduces to the $\CC$--plane one. In that limit, the solutions then behave as 
$$
f_{R}^i\sim\tau^{n_{R}^i}\,,\;\;\;\;\; f_{L}^i\sim\tau^{-n_{L}^i}\,,
$$
meaning that the only states to be considered have $n_{R}^i\ge0$ and $n_{L}^i\le0$. Clearly, $n_{L,R}^2$ must be expressed in terms of $n_{L,R}^1$, if the two wave functions $\psi_{L,R}^i$ have to describe, on the cylinder, the same fermion state. According to Eq.~(\ref{cooch}), one has
$$
\ds{n_{R}^2=-\left(n_{R}^1+\frac{\tilde\kappa_1+\tilde\kappa_2}2+1\right)}\,,\;\;\;\;\;\;\;
\ds{n_{L}^2=-\left(n_{L}^1+\frac{\tilde\kappa_1+\tilde\kappa_2}2-1\right)}\,.
$$
Imposing $n_{R}^2\ge0$, $n_{L}^2\le0$ then implies an upper bound for the allowed values of $n_{R}^1$ and a lower one for $n_{L}^1$. Therefore, it is
\be
\ds{0\le n_{R}^1\le -\frac{\tilde\kappa_1+\tilde\kappa_2}2-1}\,,\;\;\;\;\;\;\;
\ds{-\frac{\tilde\kappa_1+\tilde\kappa_2}2+1\le n_{L}^1\le0}\,,
\label{bond}
\ee
so that $|(\tilde\kappa_1+\tilde\kappa_2)/2|$ zero modes, $L$--handed if $(\tilde\kappa_1+\tilde\kappa_2)/2$ is positive, $R$--handed if it is negative, are found on ${\mc C}$. This is in agreement with what results from applying the Atiyah--Singer index theorem. Besides of the chirality and number of the zero-modes, their wave function on the cylinder ({\it i.e.} in the bulk) can also be computed, regardless to the detailed profile of ${\mc C}$, since the bulk equation is universal. In the $i=1$ coordinate system, for $\tau_1\ge\eta$, the solutions have the form
\be
\left\{
\begin{array}{l}
f_{R}^1(\tau_1)=A_R\ds{e^{\frac{\lambda_{R}^1}{r}\tau_1}}\\
f_{L}^1(\tau_1)=A_L\ds{e^{-\frac{\lambda_{L}^1}{r}\tau_1}}
\end{array}\right.\,,
\label{bulks}
\ee
where $\lambda_{R}^1=n_{R}^1+\frac12(\tilde\kappa_1+1)$, $\lambda_{L}^1=n_{L}^1+\frac12(\tilde\kappa_1-1)$. The solutions in the $i=2$ coordinates are easily obtained by transforming Eq.~(\ref{bulks}) according to Eq.~(\ref{cooch}). The bulk profile of the $R$-handed states, then, is localized at $\tau_1=\eta$ (the "$1$" fixed point) if $\lambda_{R}^1<0$, at $\tau_1=L-\eta$ (the "$2$" fixed point) if $\lambda_{R}^1>0$. The contrary holds  for the $L$-handed ones. The usual constant orbifold bulk zero mode is obtained for $\lambda_{L,R}^1=0$.

The fermion spectrum on ${\mc C}$ is summarized in Table~1, and agrees with what found in \cite{io} where ${\mc C}$ was assumed to be composed by two halves--spheres connected by a cylinder. It has been re-derived here without making any reference to the detailed profile of the resolving space. Note that the presence of one ($L$-- or $R$--handed) bulk zero mode is not automatic. It only arises when the two gauge fluxes $\tilde\kappa_{1,2}$ have the same sign, its chirality depending on this sign. This case is considered in the upper part of Table~1 and corresponds, in the orbifold limit, to a model on $S^1/\Z_2$ in which the $L$--handed component of the spinor is taken to be even (if $\tilde\kappa_{1,2}>0$), or  odd (if $\tilde\kappa_{1,2}<0$). In the other case considered in the table, on the contrary, no bulk zero--mode is present. It corresponds then to fermions which are antiperiodic on the $S^1$ circle. This is the same as considering a fermion on the segment with, at the "$1$" extremal, Neumann ($\partial \phi=0$) and Dirichlet ($\phi=0$) boundary conditions, respectively, for the $L$-- and $R$--handed components; the opposite at "$2$". The case $\tilde\kappa_{1}<0$, $\tilde\kappa_{1}>0$  is obtained from Table~2 by interchanging the two fixed points. These correspondences, which can be checked here at the level of bulk zero--modes only, will be verified in the following section, when bulk massive states will be studied. The mass--spectrum and the wave--functions will be shown to reproduce, once the orbifold limit is taken, the ones on the segment with the appropriate boundary conditions.
\begin{table}[t]
\begin{center}
\begin{tabular}{c}
\begin{tabular}{|c|c|}
\hline
$\tilde\kappa_{1,2}>0$ & $\tilde\kappa_{1,2}<0$ \\
\hline
$(|\tilde\kappa_{1}|-1)/2$ $L$ at "$1$" & $(|\tilde\kappa_{1}|-1)/2$ $R$ at "$1$"\\
\hline
$1$ $L$ in the bulk & $1$ $R$ in the bulk\\
\hline
$(|\tilde\kappa_{2}|-1)/2$ $L$ at "$2$" & $(|\tilde\kappa_{2}|-1)/2$ $R$ at "$2$"
\\ \hline
\end{tabular}
\vspace{0.1cm}\\
\begin{tabular}{|ccc|}
\hline
&$\tilde\kappa_{1}>0$, $\tilde\kappa_{2}<0$&\\
\hline
$\tilde\kappa_{1}+\tilde\kappa_{2}>0$ \hfill\vline\hfill&$\tilde\kappa_{1}+\tilde\kappa_{2}=0$ \hfill\vline\hfill& $\tilde\kappa_{1}+\tilde\kappa_{2}<0$\\
\hline
$|\tilde\kappa_{1}+\tilde\kappa_{2}|/2$ $L$ at "$1$" \hfill\vline\hfill&
 \hfill\vline\hfill& $|\tilde\kappa_{1}+\tilde\kappa_{2}|/2$ $R$ at "$2$"\\
 \hline
\end{tabular}
\end{tabular}
\caption{\footnotesize{Fermion zero--modes spectrum on the resolved $S^1/\Z_2$ orbifold for different choices of the gauge fluxes ${\tilde\kappa}_{1,2}$. The chirality and the location of the states are indicated.}}
\end{center}
\end{table}

\subsubsection{Massive states}

With the ansatz (\ref{ancim}), the Dirac equation becomes
\be
\left\{\begin{array}{l}
\partial_{\tau_i}(if_{L}^i)+\frac 1{\rho_i(\tau_i)}\left[n_i+1+(A_{\phi_i}^i-\frac12\omega_{\phi_i}^i)\right](if_{L}^i)
=mf_{R}^i\\
\partial_{\tau_i} f_{R}^i-\frac 1{\rho_i(\tau_i)}\left[n_i+(A_{\phi_i}^i+\frac12\omega_{\phi_i}^i)\right]f_{R}^i=-m(if_{L}^i)
\end{array}\right.\,,
\label{eqmcig}
\ee
and, clearly, cannot be solved until the shape of $\rho_i(\tau_i)$ and $A_{\phi_i}(\tau_i)$ is not specified. At $\tau_i\sim0$, however, it simply reduces to the $\CC$--plane one and its regular solutions can be expressed as 
\be
f_{R}^i= {\mc N}_i\,J_n(m\tau_i)\,,\;\;\;\;\;f_{L}^i= {\mc N}_i\,J_{n+1}(m\tau_i)\,,
\label{zerobecig}
\ee
for $\tau_i\ll\eta$. At $\tau_i\ge\eta$, on the contrary, the equations become those on the cylinder, and the solution assumes the form
\be
\begin{array}{l}
f_{R}^i=\alpha_i e^{i\omega_i\tau_i}+\beta_i e^{-i\omega\tau_i}\,,\\
if_{L}^i=\alpha_i \left(\frac{\lambda_i}{mr}-\frac{i\omega_i}m\right)e^{i\omega_i\tau_i}+
\beta_i \left(\frac{\lambda_i}{mr}+\frac{i\omega_i}m\right)e^{-i\omega_i\tau_i}\,,
\end{array}
\label{etasol}
\ee
having defined the integers $\lambda_i=n_i+\frac{(\tilde\kappa_i+1)}2$ and $\omega_{i}^2=m^2-{\lambda_i}^2/r^2$, which can be either positive or negative, having neglected the case $\omega_i=0$. The coefficients $\alpha_{1,2}$ and $\beta_{1,2}$ entering in Eq.~(\ref{etasol}) are determined by the evolution of the solution, due to Eq.~(\ref{eqmcig}), from the initial condition (\ref{zerobecig}) at $\tau_i=0$, to $\tau_i=\eta$. They are then fixed by the ${\mc N}_{1,2}$ coefficients appearing in Eq.~(\ref{zerobecig}). A profile--independent approach to the computation of the wave functions cannot be followed and a definite class of "trial" profiles must be used. The ${\mc C}^\infty$ profiles of Eq.~(\ref{cinprs1}), labelled with the real parameter $\delta$, will be considered, and a gauge connection $A^i=\kappa_i/2\omega^i$ will be employed. The results which follow have been verified to depend weakly on $\delta$, the strict limit $\eta\rightarrow0$ being completely resolution--independent. From now on, the case  $\delta=\eta$ will be considered.
\begin{figure}[t]
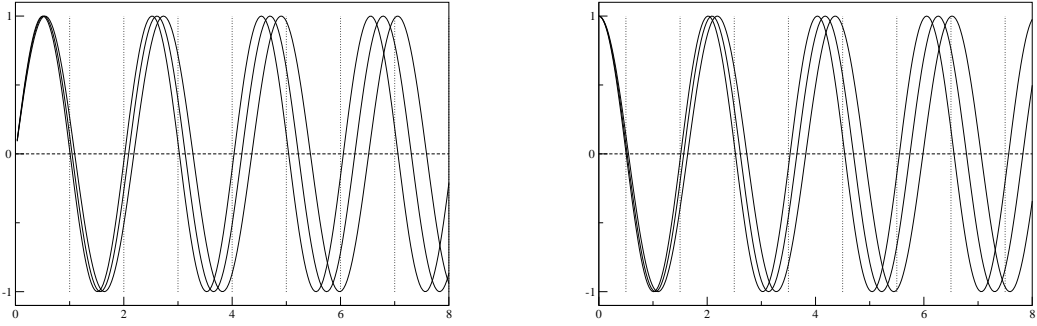

\begin{center}
\vspace{-1cm}
\includegraphics*[width=6cm]{plot_massEq.eps}
\hspace{1.5cm}
\includegraphics*[width=6cm]{plot_massEq2.eps}
\caption{\footnotesize{The function $\sigma(m)$, whose zeros give the mass--spectrum on ${\mc C}$, is plotted versus $mL/\pi$. On the left, the case $\tilde\kappa_1=1$, $\tilde\kappa_2=3$ and $n_1=1$, for three different choices of the resolution parameter $\eta=L/10,L/20,L/100$. On the right, $\tilde\kappa_1=1$, $\tilde\kappa_2=-3$ and $n_1=1$, for $\eta=L/10,L/20,L/100$}}
\end{center}
\end{figure}
Eq.~(\ref{eqmcig}) is numerically solved from $0$ to $\eta$, for any given value of the mass--parameter $m$. The solutions are found in the two coordinate sets, up to the constant multiplicative factors ${\mc N}_{1,2}$, by considering Eq.~(\ref{zerobecig}) with ${\mc N}_{1,2}=1$ as the initial condition at $\tau_i=0$. \footnote{Clearly, since the differential equation becomes singular at $\tau_i=0$, this point cannot be used for assigning the initial conditions. The initial condition (\ref{zerobecig}) has then been imposed at a point $\bar\tau_i\ll\eta$, and checks have been performed the results not to depend on $\bar\tau_i$, if it is small enough. It has been also verified, by solving the equation from $\bar\tau_i$ backwards, that the solution so obtained remains finite when $\tau_i$ approaches $0$, while divergences are encountered if perturbing the initial conditions, meaning that the regular solution of Eq.~(\ref{eqmcig}) is correctly selected by this procedure.} The parameters $\alpha_{1,2}$ and $\beta_{1,2}$ of Eq.~(\ref{etasol}) are then determined by continuity, and the two wave--functions on the cylinder are found. Since the two must describe a single spinor field on ${\mc C}$, however, they must be related by  Eq.~(\ref{cooch}), which implies $n_{1,2}$ to be related as 
\be
n_2=-n_1-1-\frac{(\tilde\kappa_1+\tilde\kappa_2)}2\,,
\label{n12}
\ee
and consequently $\lambda_2=-\lambda_1\equiv\lambda$, $\omega_1=\omega_2\equiv\omega$, but also
\be
\begin{array}{c}
f_{R}^2(L-\tau_1)\equiv\gamma f_{R}^1(\tau_1)\,,\\
if_{L}^2(L-\tau_1)\equiv-\gamma if_{L}^1(\tau_1)\,,
\end{array}
\label{masseq}
\ee
for some proportionality factor $\gamma$. Note that, once Eq.~(\ref{n12}) is imposed, the solution in each coordinate system is uniquely determined for any given $n_1$, up to the rescaling ${\mc N}_{1,2}$, which at most can change the value of $\gamma$. All what can be done is then to try to check if Eq.~(\ref{masseq}) is satisfied. To this end, define
$$
\sigma(m)=\frac1{|\vec{f}^1(L/2)||\vec{f}^2(L/2)|}\textrm{Det}\left(\begin{array}{cc}
f_{R}^1(L/2) &  -f_{R}^2(L/2)\\
if_{L}^1(L/2)     & if_{L}^2(L/2)
\end{array}\right)\,,
$$
where $|\vec{f}^{1,2}(L/2)|$ is the modulus of the $(f_{R}^{1,2},\pm if_{L}^{1,2})$ vector, so that $\sigma$ is the sine of the angle between $\vec{f}^1(L/2)$ and $\vec{f}^2(L/2)$. It only vanishes if (\ref{masseq}) is satisfied at $\tau_{1,2}=L/2$. Vanishing of $\sigma(m)$, however, is equivalent to Eq.~(\ref{masseq}), since functions in the two sides of Eq.~(\ref{masseq}) are solutions, on the cylinder, to the same differential equation and then coincide if they are equal at one point. The mass--spectrum of fermions on ${\mc C}$ is given by the values of $m$ for which $\sigma(m)$ vanishes.
\begin{figure}[t]
\begin{center}
\vspace{-1cm}
\includegraphics*[width=4cm,angle=-90]{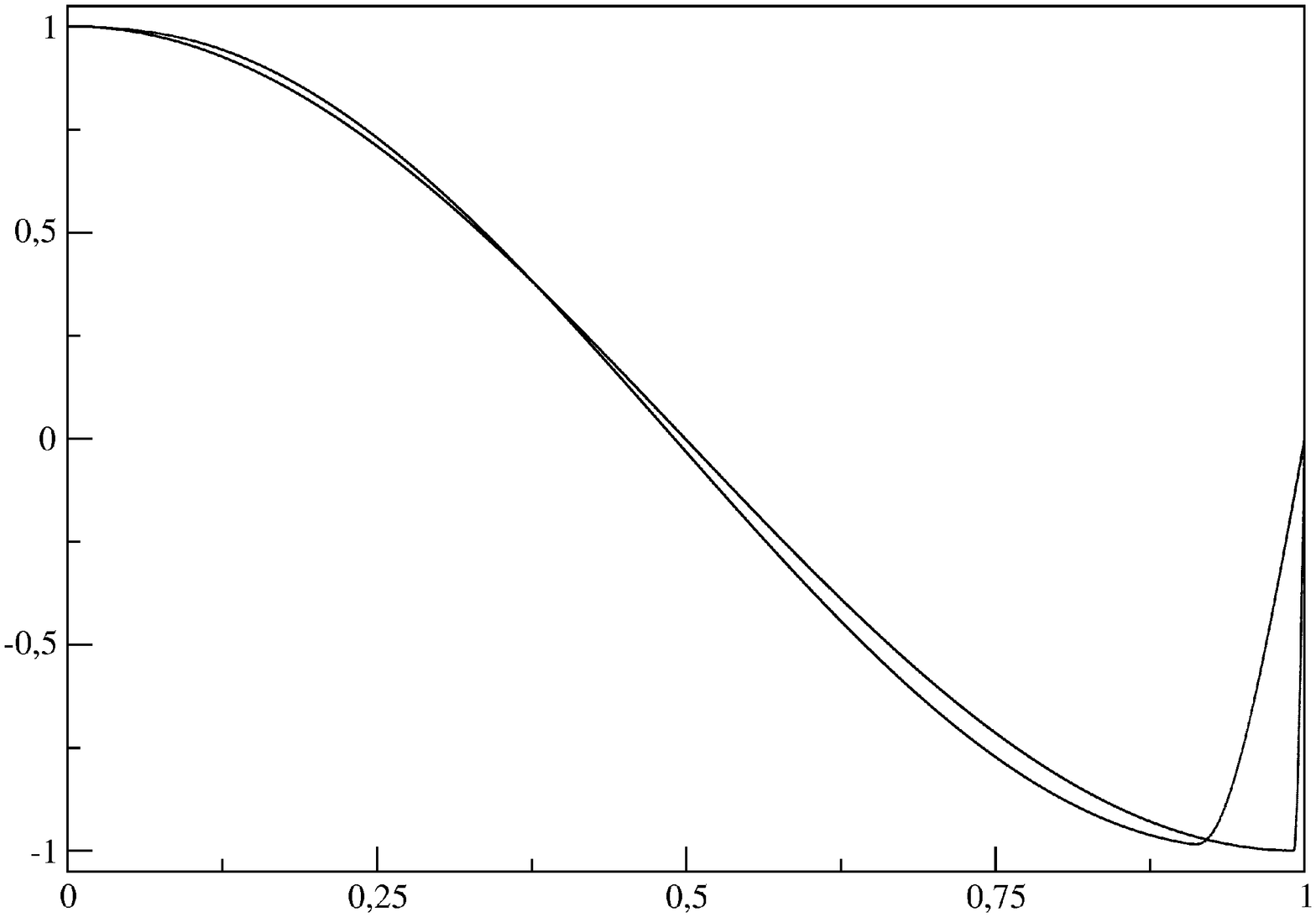}
\hspace{1cm}
\includegraphics*[width=4cm,angle=-90]{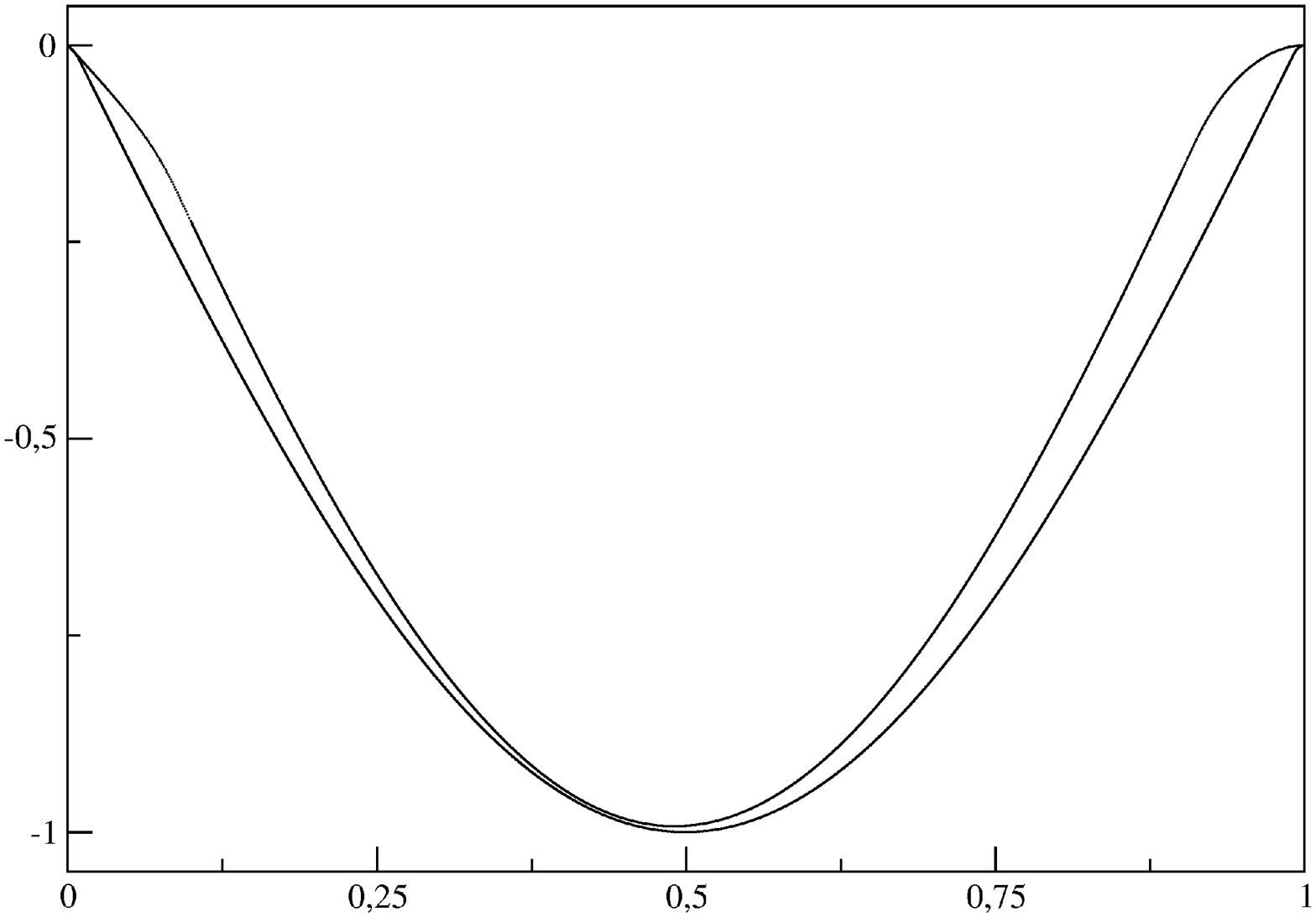}
\caption{\footnotesize{The $L$-- and $R$--handed components of the first Kaluza--Klein mode, for two different values of the resolution parameter $\eta=L/10,\,L/100$, as a function of $\tau/L\in[0,1]$. The case $\tilde\kappa_1=1$, $\tilde\kappa_2=3$ and $n_1=-1$ is considered in the picture.}}
\end{center}
\end{figure}

As an example, consider the case $\tilde\kappa_1=1$, $\tilde\kappa_2=3$. Only for $n_1=-1$, and then $\lambda=0$, small masses are obtained, since it happens that  $\sigma=0$ can be only realized when $\omega^2=m^2-{\lambda}^2/r^2$ is positive, the bulk solutions having oscillating behavior. All states with $\lambda\neq0$ have then mass at the $1/r$ scale, and decouple in the orbifold limit.
A plot of $\sigma(m)$ is shown in Fig.~1 for different choices of the resolution parameter $\eta$. The zeroes of $\sigma$ are seen to approach, as $\eta$ decreases, the expected values of $m_k=k\pi/L$. This is consistent with the interpretation of ${\mc C}$ as an $S^1/\Z_2$ orbifold with even $L$--handed fields.
Note that smaller values of the orbifold masses are better reproduced for a given resolution parameter $\eta$. For the first Kaluza-Klein state the relative errors with respect to the orbifold value $\pi/L$ are  
$(0.09,0.04,0.009)$, linearly decreasing with $\eta=L/10,L/20,L/100$. For the second Kaluza-Klein state the relative errors are $(0.09,0.05,0.008)$, and similarly for the third one: $(0.09,0.04,0.009)$. One has, in practice, $\delta m/m\sim \eta/L$. Once the values of the masses are found by computing at the zeroes of $\sigma(m)$, the wave function for each massive state is easily computed. In Fig.~2, the $L$-- and $R$--handed components of the wave function of the first Kaluza--Klein state are plotted, for $\eta=L/10,\,L/100$, as a function of $\tau\in[0,L]$. The wave--functions are given, for $\tau\in[0,\,L-\eta]$, by $f_{R}^1(\tau)$ and $if_{L}^1(\tau)$; by $f_{R}^2(L-\tau)$ and $-if_{L}^1(L-\tau)$, rescaled so that the resulting profile is continuous, in the $[L-\eta,\,L]$ interval. The $L$--handed component, as shown in the plot, approximates the cosine function with  frequency $L/\pi$, while the $R$--handed one resambles a sine with the same frequency and a minus sign in front. This is precisely what expected for the wave function of the first Kaluza--Klein state on $S^1/\Z_2$. To conclude, the case $\tilde\kappa_1=1$, $\tilde\kappa_2=-3$ can be considered. It should correspond, in the orbifold limit, to a segment with Neumann (Dirichlet) boundary condition for the $L$--($R$--)handed field at the "$1$" extremal and the contrary at "$2$". In Fig.~1, the profile of $\sigma(m)$ obtained in this case is plotted for $\eta=L/10,\,L/20,\,L/100$. The zeroes approach now, as expected, $m_k=\pi/L(k+1/2)$. One could also verify that the wave functions are correctly reproduced.

\section*{Acknowledgments}
I'm grateful to M.~Serone for the continuous support and the many useful discussions and suggestions. I also thank F.~Zwirner for reading the manuscript.

\appendix

\section{Two dimensional spaces with $O(2)$ isometry group}

\subsection{The metric tensor in polar coordinates} 
 
The $O(2)$ group divides in $SO(2)\otimes \Z_2$; it will exist a coordinate system $\{x^i\}$ on 
which $SO(2)$ acts as rotation, and $\Z_2$ as parity:
\be
\begin{array}{l}
x^i\;\;\longrightarrow\;\;(h_\lambda)^{i}_{\ j}\,x^j\,,\\
x^i\;\;\longrightarrow\;\;(P)^{i}_{\ j}\,x^j\,,
\end{array}
\label{act}
\ee
where $h_\lambda=\ds{e^{i\lambda \sigma_2}}$ and $P=\sigma_3$. The invariance under $O(2)$ of the line element implies for the metric $g_{ij}$ to satisfy
\be
\begin{array}{l}
g_{ij}(h_\lambda x)\,(h_\lambda)^{i}_{\ l}\,(h_\lambda)^{j}_{\ k}=g_{lk}(x)\,,\\
g_{ij}(P x)\,(P)^{i}_{\ l}\,(P)^{j}_{\ k}=g_{lk}(x)\,.
\end{array}
\label{iso-c}
\ee
In this coordinate system, of course, the metric must be regular (${\mc C}^\infty$, with rank $2$ and $(+,+)$ signature) at any point, including the origin $x=0$, which has however the peculiar property of being fixed under $O(2)$. Take Eq.~(\ref{iso-c}) at $x=0$, it states the $O(2)$ invariance of $g_{ij}(0)$, 
which transforms as $2$-tensor (in the ${\bf 2}\otimes {\bf 2}$) of $O(2)$. Since the only invariant $O(2)$ 
tensor is $\delta_{ij}$, $g$ must be proportional to the identity at $x=0$. 
Moreover, if $d$ derivatives of Eq.~(\ref{iso-c}) are taken, one finds that, at $x=0$, the $d$-th derivative of $g$, which is in the ${\bf 2}^{d+2}$ tensor representation, must be invariant. A non trivial invariant can be only built when $d$ is even, meaning that all odd derivatives of $g_{ij}$ are enforced to vanish at zero.

The space will be more simply described in the polar coordinates $(\tau,\,\phi)$, being $\phi$ an angle and $\tau$ running on the positive real axis. The coordinate transformation reads
\be
x^i=(h_\phi)^{i}_{\ j}\,v^j(\tau)\,,
\label{coo}
\ee
being $v(\tau)$ a representative of $\RR^2/SO(2)$ which can be chosen to have the form 
$v^i(\tau)=\left(r(\tau),\,0\right)^i$, with $r(\tau)$ a smooth positive monotonic function vanishing at $\tau=0$ and divergent for $\tau\rightarrow\infty$. Of course, this coordinate system is not appropriate for describing the origin, Eq.~(\ref{coo}) being not invertible at $x=0$, which indeed corresponds to $\tau=0$ for any value of $\phi$. In polar coordinates, the action of $h_\lambda$ is simply a shift $\phi\rightarrow\phi+\lambda$. Accordingly, the Killing vector field $K$ associated to $SO(2)$ has components $K^\tau=0$, $K^\phi=1$ and the Killing equation $\left({\mc L}_K(g)\right)_{ij}=0$ states that
\be
\partial_\phi g_{\tau\tau}=\partial_\phi g_{\tau\phi}=\partial_\phi g_{\phi\phi}=0\,,
\label{fi-i}
\ee
meaning that, in polar coordinates, all the components of the metric only depend on $\tau$. Moreover, since the parity $P\in O(2)$ acts as $\phi\rightarrow-\phi$ in polar coordinates, $g_{\tau\phi}$ must vanish. The metric, up to now, is parametrized in terms of the two functions of $\tau$, $g_{\tau\tau}$ and $g_{\phi\phi}$. The possibility of choosing $r(\tau)$, however, can be used to fix $g_{\tau\tau}$ to $1$. This can be implemented by taking
\be
\tau(r)=\int_{0}^r dr'\sqrt{g_{11}\left[x=(r',0)\right]}\,,
\label{taur}
\ee
which is indeed monotonic, invertible and positive. 

Summarizing, the line element has the general form
\be
ds^2=d\tau^2+\rho^2(\tau)d\phi^2\,,
\label{ls}
\ee
having defined the function
\be
\rho(\tau)=\sqrt{g_{\phi\phi}(\tau)}=r(\tau)\sqrt{g_{22}\left[x=(r(\tau),0)\right]}\,,
\label{rho}
\ee
which is, of course, undetermined, and encloses all the freedom one has in the definition of ${\mc R}$. Note that $\rho(\tau)$ is almost arbitrary, but not completely. At $\tau=0$, in particular, it must satisfy some consistency conditions which ensure that the space, described  in a coordinate system which is ill-defined in the origin, is however regular at that point, and the singularity is entirely due to the coordinate choice. By taking $\partial_\tau$ derivatives of Eq.~(\ref{rho}), and remembering that odd derivatives of $g_{ij}$ vanish at zero, one easily realizes that
\be
\rho(0)=0\,,\;\;\;\;\;\frac{d\rho}{d\tau}(\tau=0)=1\,,\;\;\;\;\;\frac{d^{2n}}{d\tau^{2n}}\rho(\tau=0)=0\,,
\label{con}
\ee
for any $n$ positive integer.

\subsection{Globally defined one-forms}

Be $\Omega$ a globally defined $1$-form field, invariant under the isometry group $O(2)$, with negative intrinsic parity under the $\Z_2\subset O(2)$. Its components $\Omega_i(x)$ in the "$x$" coordinate system considered in the previous section, in which $O(2)$ acts as in Eq.~(\ref{act}) are subjected to the condition
\be
\begin{array}{l}
\Omega_i(h_\lambda x)(h_\lambda)^{i}_{\ j}=\Omega_j(x)\,,\\
\Omega_i(P x)(P)^{i}_{\ j}=-\,\Omega_j(x)\,,
\end{array}
\label{in}
\ee
By looking at Eq.~(\ref{in}) and its derivatives at $x=0$, one immediately recognizes that any $d$-th derivative of $\Omega_i$ must vanish at the origin if $d$ is even, since it transforms in the ${\bf 2}^{d+1}$ of $O(2)$. The Killing equation $\left({\mc L}_K(\Omega)\right)_i=0$, with $K$ as in the previous section, states that both $\phi$ and $\tau$ components of $\Omega$ only depend on $\tau$. Moreover, parity invariance $\phi\rightarrow-\phi$ enforces $\Omega_\tau(\tau)$ to be identically zero. The more general $O(2)$ invariant vector field $\Omega$ is then parametrized in term of one function of $\tau$ only, its $\phi$ component $\Omega_\phi(\tau)$, which can be expressed as
$$
\Omega_\phi(\tau)=\Omega(\partial_\phi)=-r(\tau)\Omega_2\left[x=(r(\tau),0)\right]\,,
$$
in terms of the components $\Omega_i(x)$ of $\Omega$ in the "$x$" coordinates. Taking derivatives of the above equation w.r.t. $\tau$ and remembering that $\Omega_i$ and its even derivatives need to vanish at the origin, one immediately recognizes that $\Omega_\phi(\tau)$ must satisfy the following consistency conditions at $\tau=0$
\be
\Omega_\phi(\tau=0)=0\,,\;\;\;\;\;\;\frac{d^{n}}{d\tau^n}\Omega_\phi(\tau=0)=0\,,
\label{00}
\ee
where $n$ is any odd positive number. Conditions (\ref{00}), analogously to those in Eq.~(\ref{con}), are necessaries to ensure $\Omega$ to be well defined at the origin.

\section{Massless fermions on $T^2/\Z_N$}

A resolved version ${\mc R}_N$ of the $T^2/\Z_N$ orbifold will be considered, and the massless Dirac equation studied on it. The resulting bulk--brane field distributions will be derived.

\subsection{The resolving space}

The $T^2/\Z_N$ orbifolds are defined by identifying points on the torus $T^2$ which are related by the $2\pi/N$ Lorentz rotation 
\be
z\rightarrow\tau z\,.
\label{rot}
\ee
The covering $T^2$ is obtained from the $\CC$ plane with Euclidean metric by the identification
\be
z\sim z+n+mU\,,
\label{tra}
\ee
where the complex number $U$ ($\textrm{Im}(U)\neq0$) is the complex structure, $m$ and $n$ are arbitrary integers. Not for any $N$ and $U$, however, the Lorentz rotation (\ref{rot}) is compatible with the $T^2$ identification (\ref{tra}). The only possibilities, shown in Fig.~3, are $N=2$, for any value of $U$, and $N=3,4,6$, in which cases the complex structure $U$ must be equal to the $\Z_N$ phase $\tau=e^{2\pi i/N}$. A spinor field $\psi$ on $T^2/\Z_N$ is defined as a field on $\CC$ which remains invariant, up to a $U(1)$ phase transformation, under the torus lattice translations
\be
\psi(z+1)=T_1\psi(z)\,,\;\;\;\;\;\psi(z+U)=T_U\psi(z)\,,\;\;\;\;\; T_{1,U}\equiv{\ds e^{\frac{2\pi i}{N}t_{1,U}} }\,,
\label{tras}
\ee
and orbifold $2\pi/N$ rotations
\be
\psi(\tau z)={\mc P}\psi(z)\,,\;\;\;\;\;\;\;\;{\mc P}\equiv{\ds e^{\pi i\left(1-\frac1N\right){\sigma_3}}
e^{\frac{\pi i}{N}p} }\,.
\label{rots}
\ee
Consistency with the geometric action of translations and rotations requires constraints on the allowed values of $t_{1,U}$ and $p$ in Eq.~(\ref{tras}) and (\ref{rots}) (see {\it i.e.} \cite{?}). For $T^2/\Z_2$ one needs $p=\pm1$, $t_{1,U}=0,1$. For $T^2/\Z_3$, $p$ is in the $-2,0,2$ range while $t_{1}=t_U=0,1,2$. In the $T^2/\Z_4$ case, one has $p=\pm3,\pm1$ and $t_1=t_U=0,1$. For $T^2/\Z_6$, finally, $t_1=t_U=0$ and $p=\pm5,\pm3,\pm1$. The fixed points of $T^2/\Z_N$ are labelled as $z_{l,i}$ ($i=1,\ldots,N_l$), where the integer $l$, which assumes values in the $1,\ldots,\, [N/2]$ range, is the minimum power of $\tau$ for which
$$
 z_{l,i}=\tau^l z_{l,i}+m_{i}+n_iU\,,
$$
for some couple of integers $(m_i,n_i)$. In the following, a point $z_{l,i}$ will be called an "$l$--fixed point". On the fundamental domain of $T^2/\Z_2$, shown in Fig.~3, four distinct $1$--fixed points are present, with $(n_i,m_i)=(0,0),\,(1,0),\,(1,1),\,(0,1)$ for $i=1,\ldots,4$, respectively. On $T^2/\Z_3$, one has three $1$--fixed points with $(n_i,m_i)=(0,0),\,(1,0),\,(1,1)$  while two $1$--fixed  ($(n_i,m_i)=(0,0),\,(1,0)$) and one $2$--fixed ($(n,m)=(1,0)$) points are present on $T^2/\Z_4$. Finally, $T^2/\Z_6$ has one $1$--fixed $(0,0)$, one $2$--fixed $(1,0)$ and one $3$--fixed $(0,1)$ point.
\begin{figure}[t]
\begin{center}
\begin{picture}(360,120)(0,-10)
\put(45,72){\begin{picture}(110,85)(0,0)
\GBox(25,0)(55,60){0.8}
\Vertex(25,0){3}
\Vertex(55,30){3}
\Vertex(55,0){3}
\Vertex(25,30){3}
\Line(25,0)(85,0)
\Line(85,0)(85,60)
\Line(85,60)(25,60)
\Line(25,60)(25,0)
\ArrowLine(55,0)(55,30)
\ArrowLine(25,0)(25,30)
\ArrowLine(55,60)(25,60)
\ArrowLine(55,0)(25,0)
\ArrowLine(55,60)(55,30)
\ArrowLine(25,60)(25,30)
\Text(20,15)[c]{\footnotesize $A$}
\Text(20,45)[c]{\footnotesize $B$}
\Text(60,45)[c]{\footnotesize $D$}
\Text(60,17)[c]{\footnotesize $C$}
\Text(40,-6)[c]{\footnotesize $E$}
\Text(40,66)[c]{\footnotesize $F$}
\end{picture}}
\put(45,-11){\begin{picture}(110,85)(0,0)
\GBox(25,0)(55,30){0.8}
\Vertex(25,0){3}
\Vertex(55,30){3}
\Vertex(55,0){2}
\Line(25,0)(85,0)
\Line(85,0)(85,60)
\Line(85,60)(25,60)
\Line(25,60)(25,0)
\ArrowLine(55,0)(55,30)
\ArrowLine(25,30)(25,0)
\ArrowLine(25,30)(55,30)
\ArrowLine(55,0)(25,0)
\Text(20,15)[c]{\footnotesize $A$}
\Text(40,35)[c]{\footnotesize $B$}
\Text(60,17)[c]{\footnotesize $C$}
\Text(40,-6)[c]{\footnotesize $D$}
\end{picture}}
\put(205,-11){\begin{picture}(110,85)(0,0)
\SetColor{White}
\GBox(0,0)(36.67,63.51){0.8}
\begin{rotate}{30}
\BBoxc(25,-10)(50,20)
\begin{rotate}{30}
\BBoxc(45,19)(90,37)
\end{rotate}
\end{rotate}
\SetColor{Black}
\Vertex(0,0){3}
\Line(0,0)(73.34,0)
\Line(73.34,0)(110.01,63.51)
\Line(110.01,63.51)(36.67,63.51)
\Line(36.67,63.51)(0,0)
\Vertex(36.67,21.17){2}
\Vertex(18.33,31.76){1}
\ArrowLine(0,0)(36.67,21.17)
\ArrowLine(0,0)(18.33,31.76)
\ArrowLine(36.67,63.51)(18.33,31.76)
\ArrowLine(36.67,63.51)(36.67,21.17)
\Text(2,19)[c]{\footnotesize $A$}
\Text(24,52)[c]{\footnotesize $D$}
\Text(43,44)[c]{\footnotesize $C$}
\Text(22,7)[c]{\footnotesize $B$}
\end{picture}}
\put(205,72){\begin{picture}(110,85)(0,0)
\SetColor{White}
\GBox(36.67,0)(73.34,63.51){0.8}
\begin{turn}{30}
\BBoxc(70,30)(50,23)
\BBoxc(60,-30)(50,23)
\end{turn}
\SetColor{Black}
\Vertex(36.67,0){3}
\Line(36.67,0)(110,0)
\Line(36.67,0)(0,63.51)
\Line(0,63.51)(73.34,63.51)
\Line(73.34,63.51)(110,0)
\Vertex(36.67,42.34){3}
\Vertex(73.33,21.17){3}
\ArrowLine(36.67,0)(36.67,42.34)
\ArrowLine(36.67,0)(73.33,21.17)
\ArrowLine(73.34,63.51)(36.67,42.34)
\ArrowLine(73.34,63.51)(73.33,21.17)
\Text(31,23)[c]{\footnotesize $A$}
\Text(51,56)[c]{\footnotesize $D$}
\Text(79,42)[c]{\footnotesize $C$}
\Text(60,8)[c]{\footnotesize $B$}
\end{picture}}
\end{picture}
\caption{
\footnotesize The picture shows the $T^2/\Z_2$ $T^2/\Z_3$, $T^2/\Z_4$
and $T^2/\Z_6$ orbifolds and their covering tori. Points of decreasing size indicate the $1$--, $2$-- and $3$--fixed points respectively. The Grey region represents the fundamental domain of the
orbifolds, and the segments delimiting it must be identified according to:
$A\sim D$, $B\sim C$ and, in the $T^2/\Z_2$ case, $E\sim F$.
}
\end{center}
\end{figure}
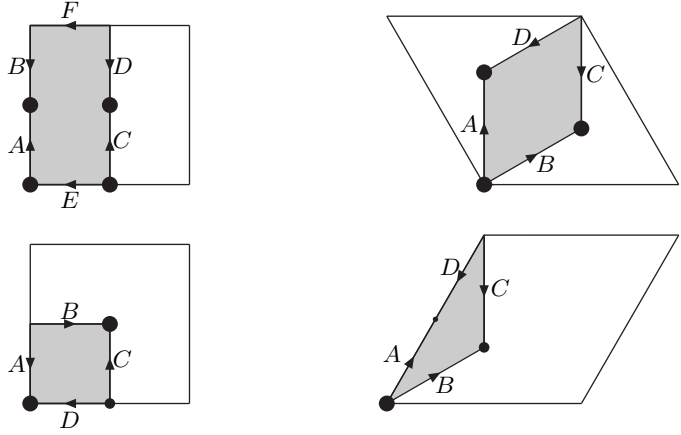
At each fixed point $z_{l,i}$, the orbifold has a conical $\CC/\Z_{N/l}$ singularity; the twist matrix which characterizes it is the effective orbifold twist ${\mc P}_{l,i}$
$$
\begin{array}{r}
\psi(z_{l,i}+\tau^l z)=\psi(\tau^l(z_{l,i}+z)+m_i+n_i U)\\\vspace{-4mm}\\
=T_{1}^{m_i}T_{U}^{n_i}{\mc P}^l\psi(z_{l,i}+ z)\equiv
{\mc P}_{l,i}\psi(z_{l,i}+ z)\,,
\end{array}
$$
{\it i.e.} the transformation matrix of orbifold fields under $2\pi l/N$ rotations around $z_{l,i}$. Clearly, ${{\mc P}_{l,i}}^{N/l}=1$ by consistency.

A resolution ${\mc R}_N$ of each $T^2/\Z_N$ orbifold is easily constructed by removing small disks surrounding the singularities ---which correspond, for each $z_{l,i}$ fixed point, to a truncated $\CC/\Z_{N/l}$ orbifold with projection matrix ${\mc P}_{l,i}$--- from the fundamental domain and replacing them with the appropriate resolving space, defined in Sect.~3. In this way, the resolution simply consists on a flat region, corresponding to the bulk, which connects various resolved cones that represent the singularities. Note that, since ${{\mc P}_{l,i}}^{N/l}=1$, the projection matrices can be written as 
$$
{\mc P}_{l,i}={\ds e^{\pi i\left(1-\frac{l}N\right){\sigma_3}}
e^{\frac{\pi i}{N/l}p_{l,i}} }\,,
$$
with $p_{l,i}$ running from $-(N/l-1)$ to $N/l-1$ at steps of two. This matches the form (\ref{bcN}) of the projection matrix, so that the results of Sect.~3 can be directly applied if identifying $N$ and $p$ with $N/l$ and $p_{l,i}$, respectively. Clearly, as in Eq.~(\ref{kap}), an arbitrary integer monopole charge $q_{l,i}$ can be put at each fixed point. As shown in Sect.~4.1 and 4.2, when $|p|$ is maximal the sign of $q$ needs to be the same as $p$ for the resolution correctly reproduce the $\CC/\Z_N$ orbifold. For each orbifold, the allowed configurations of effective projections $p_{l,i}$ are easily derived. In the $T^2/\Z_2$ case either, for $t_{1,U}=0$, all fixed points have the same twist $p=\pm 1$ or, if at least one of $t_{1,U}$ is different from zero, two fixed points have the opposite twist than the other two. For $T^2/\Z_3$, similarly, either the twist for all fixed points is the same, $p_{1,i}=p=\pm2,0$, or they are all different, $p_{i,i}=(2,0,-2)$ or permutations. In the $T^2/\Z_4$ case one may have $p_{1,2}=p_{1,1}=p=\pm3,\pm1$ and ${\mc P}_{2,1}={\mc P}^2$, and then $p_{2,1}=p+2\, \textrm{Mod}(4)=(-1)^{(p+1)/2}$, or $p_{1,2}=-p_{1,1}=-p=\pm3,\pm1$, and $p_{2,1}=p\,\textrm{Mod}(4)=(-1)^{(p-1)/2}$. For $T^2/\Z_6$, the only possibility is $p_{1,1}=p=\pm5,\pm3,\pm1$, $p_{2,1}=p+3\,\textrm{Mod}(6)$ and $p_{1,1}=p\,\textrm{Mod}(4)=(-1)^{(p-1)/2}$. In the following, the fermion zero--modes spectrum on the resolving space ${\mc R}_N$ will be derived. The index of the Dirac operator, {\it i.e.} the number of $L$-- minus the number of $R$--handed zero--modes, can be easily computed from the Atiyah--Singer index theorem. By means of Eq.~(\ref{kap}), one finds a Dirac index
\be
{\mc I}\equiv n_L-n_R=\frac1{2\pi}\int F=\sum_{l,i}\frac{p_{l,i}}{2N/l}+\sum_{l,i}q_{l,i}\,,
\label{ind}
\ee
which is integer, for any allowed choice of $p_{l,i}$. Is worth noticing that the first term in Eq.~(\ref{ind}) counts the orbifold bulk zero--modes, meaning that it is always zero but for $p=\pm(N-1)$ and $t_{1,U}=0$, in which case its value is $\pm1$, in accordance with the orbifold projection (\ref{rots}) which leaves untwisted, respectively, the $L$--($R$--)handed component of the fermion field.

\subsection{The zero--modes}

The fundamental domain of  $T^2/\Z_N$ ---from which infinitesimal disks have been removed in correspondence with the singularities--- constitutes the "bulk" of the resolving space ${\mc R}_N$ defined in the previous section.
The massless Dirac equation in this region simply states (see Eq.~(\ref{eqC})) the wave functions to be holomorphic ($\psi_R(z)$) and anti--holomorphic ($\psi_L(\bar z)$). Define the circuit $\Gamma$, counter--clockwise oriented, as the boundary of the bulk region. It is composed by infinitesimal circular paths $\gamma_{l,i}$ around each $z_{l,i}$ fixed point, and completed by the segments, indicated in Fig.~3 as $A\sim D$, $B\sim C$ (and $E\sim F$ in the $T^2/\Z_2$ case), which delimit the fundamental domain. Note that each segment is covered by $\Gamma$ in the opposite direction than its "mirror", with which it is identified, and chiral fields at the two identified segments are proportional, as implied by Eq.~(\ref{tras}) and (\ref{rots}). One then has
$$
\frac1{2\pi i}\int_{A_2-A_1}dz\partial_z\log{(\psi_R(z))}=
-\frac1{2\pi i}\int_{A_2-A_1}d\bar z\partial_{\bar z}\log{(\psi_L(\bar z))}=0\,,
$$
for $(A_1,A_2)=(A,D),\,(B,C),\,(E,F)$. Therefore, if the above contour integral is performed on the whole closed path $\Gamma$, it only receives contributions from the $\gamma_{l,i}$ circular paths, which are all oriented clockwise and cover each a $2\pi/(N/l)$ plane angle. \footnote{In the case in which two representatives of the same fixed point $z_{l,i}$ are present as two distinct corners of the fundamental domain, $\gamma_{l,i}$ must be thought as the union of two paths, each covering a $\pi/(N/l)$ angle, located at the two corners.} At each fixed point $z_{l,i}$, the $L$-- and $R$--handed wave functions behave as
\be
\psi_R(z)\sim (z-z_{l,i})^{\lambda_{R}^{l,i}}\,,\;\;\;\;\;\psi_L(\bar z)\sim (\bar z-{\bar z}_{l,i})^{-\lambda_{L}^{l,i}}\,,
\label{libe}
\ee
so that (see Eq.~(\ref{wfr})) locally match the massless wave functions ---with angular momenta $\lambda_{R,L}=\lambda_{R,L}^{l,i}$--- on the resolved $\CC/\Z_{(N/l)}$. The allowed values for $\lambda_{R,L}^{l,i}$ are then seen from Eq.~(\ref{lamLR}) to be
\be
\begin{array}{l}
\lambda_{R}^{l,i}=\frac{p_{l,i}+N/l-1}2+N/l(n_{R}^{l,i}+q_{l,i})   \,,\\
\lambda_{L}^{l,i}=\frac{p_{l,i}-(N/l-1)}2+N/l(n_{L}^{l,i}+q_{l,i})  \,,
\end{array}
\label{lamLRli}
\ee
with $n_{R,L}^{l,i}$, respectively, arbitrary positive and negative integers. By mean of Eq.~(\ref{libe}), the contour integral on $\gamma_{l,i}$ is immediately computed, and one finds
\be
\begin{array}{c}
\frac1{2\pi i}\oint_{\Gamma}dz\partial_z\log{(\psi_R(z))}=
\frac1{2\pi i}\sum_{l,i}\int_{\gamma_{l,i}}dz\left(\frac{\psi_{R}'(z)}{\psi_R(z)}\right)=
-\sum_{l,i}\frac{\lambda_{R}^{l,i}}{N/l}=b_R\,,\\
-\frac1{2\pi i}\oint_{\Gamma}d\bar z\partial_{\bar z}\log{(\psi_L(\bar z))}=
-\frac1{2\pi i}\sum_{l,i}\int_{\gamma_{l,i}}d\bar z\left(\frac{\psi_{L}'(\bar z)}{\psi_L(\bar z)}\right)=
\sum_{l,i}\frac{\lambda_{L}^{l,i}}{N/l}=b_L\,,
\end{array}
\label{cglo1}
\ee
where the theorem of residuals has been used to write the last equality. Having assumed the wave functions not to have poles in the bulk, the result is entirely given by the number $b_{R,L}$ of zeroes, counted with their multiplicity, of $\psi_{R,L}$ inside $\Gamma$. By means of Eq.~(\ref{lamLRli}), the condition (\ref{cglo1}) can be expressed as
\be
\sum_{l,i}n_{R}^{l,i}=-b_R-{\mc I}-1\,,\;\;\;\;\; \sum_{l,i}n_{L}^{l,i}=b_L-{\mc I}+1\,,
\label{cglo}
\ee
with ${\mc I}$ as in Eq.~(\ref{ind}). In the following, $b_{L,R}=0$ will be assumed, and the assumption will be verified at the end, when the number of chiral zero--modes derived with $b_{L,R}=0$ will be shown to be consistent with the known value of the index ${\mc I}$. If an additional (say, $R$--handed) state with $b_R>0$ has to be present, since all $n_{R}^{l,i}$'s need to be positives, ${\mc I}$ would need to be smaller than $-1-b_R$, which is incompatible with the existence of any "new" $L$--handed state with $b_L>0$, whose presence is however  required by the index theorem. Consider then Eq.~(\ref{cglo}) with $b_{L,R}=0$. It implies that only $L$--($R$--)handed states can be present on ${\mc R}_N$ if ${\mc I}$ is positive (negative). No $R$--handed and ${\mc I}$ $L$--handed states have then to appear if ${\mc I}$ is positive, no $L$--handed and $-{\mc I}$ $R$--handed if it is negative. The distinct solutions to Eq.~(\ref{cglo}), however, are much more than $|{\mc I}|$, meaning that not all of them are independent. One should be able to write down, for any solution to Eq.~(\ref{cglo}), the explicit form of the wave--function, identified by its behavior (\ref{libe}) at the fixed points, and check that, for any given value of ${\mc I}$, only $|{\mc I}|$ of them are linearly independent. This can be done in some explicit example for the $T^2/\Z_2$ case, the wave--functions being provided by suitable products and ratios of Jacobi theta functions, as discussed in \cite{io}. Even though a mathematical proof is not available, the general criterion is that linearly independent wave--functions are only labelled by the degree and location of their pole of maximum degree. This is physically very reasonable since, in the orbifold limit, the wave function localizes at its maximum pole, and "local" quantities only, such as the angular momentum $\lambda_{L,R}^{l,i}$, can be relevant to label it. By using the above prescription, the location and number of zero--modes can be derived, for any $T^2/\Z_N$ orbifold and any allowed value of the twists $p_{l,i}$.

As an illustrative example of the procedure, consider the simple $T^2/\Z_2$ case, with $p_{1,i}=+1$ for any $1=1,\ldots\,4$. One has ${\mc I}=1+\sum_{i=1}^4q_{1,i}>0$, since all monopole charges must be positive, and then $L$--handed state only can appear. In this case Eq.~(\ref{lamLRli}) reads $\lambda_{L}^{1,i}=2(n_{L}^{1,i}+q_{1,i})$ while Eq.~(\ref{cglo}) becomes
\be
\sum_{i}(n_{L}^{1,i}+q_{1,i})=0\,.
\label{cglo2}
\ee
Consider each fixed point separately; take $i=1$ for definiteness. A state localized at $i=1$ is found for any choice of $n_{L}^{1,i}\le0$, satisfying Eq.~(\ref{cglo2}), such that $\lambda_{L}^{1,1}>0$, all the other $\lambda_{L}^{1,i}$'s being smaller than that. Note that, having fixed a positive value for $\lambda_{L}^{1,1}$, several choices of $n_{L}^{2,3,4}$ satisfying the above constraints could be possible. Having assumed the physical states to be only labelled by the location and degree of their higher pole, however, just one must be counted among all these, so all what matters is that at least one of them exists. This is clearly the case for any $n_{L}^{1,1}=0,\ldots,-q_{1,1}+1$ ---such that $\lambda_{L}^{1,1}>0$--- since one could take  $\lambda_{L}^{1,i}=0$ for $i=2,3$ and $\lambda_{L}^{1,4}=-\lambda_{L}^{1,1}$. Therefore, $q_{1,1}$ $L$--handed localized states have been found at $i=1$ and, since the same holds for any $i$, $q_{1,i}$ states are found at each fixed point. One independent state only, the bulk one ($\lambda_{L}^{1,i}=0$) escaped the above analysis. Note that, even if this does not occur in the present case, states which are \emph{doubly localized} at two fixed points "$1$" and "$2$" can appear. States with equal angular momenta at two points $\lambda_{L}^{1,2}=\lambda_L$ are, in general, linear combination of two states, one localized at "$1$", the other at "$2$", with maximum poles of order $\lambda_L$. It happens, however, that such singly localized states does not exist, for some value of $\lambda_{L}$. When this is the case, a new linearly independent doubly localized zero--mode must be take into account.
\begin{table*}[p]
\begin{center}
\begin{tabular}{l}
\begin{tabular}{|c|c|c|c|c|}
\hline \ & $1$ & $2$ & $3$ & $4$  \\
\hline
($1,1,1,1$) & $|q_{1,1}|$ & $|q_{1,2}|$ & $|q_{1,3}|$ & $|q_{1,4}|$  \\\hline
($1,1,-1,-1$)& $\left[\frac{{\mc I}+|q_{1,1}|-|q_{1,2}|-1}2\right]_-$ & 
$\left[\frac{{\mc I}+|q_{1,2}|-|q_{1,1}|-1}2\right]_-$ & $0$ & $0$  \\ 
\hline
\end{tabular}
\vspace{0.5cm}\\
\begin{tabular}{|c|c|c|c|}
\hline 
($2,2,2$)& $1$ & $2$ & $3$  \\
\hline
 \ & $|q_{1,1}|$ & $|q_{1,2}|$ & $|q_{1,3}|$  \\\hline
\end{tabular}\vspace{0.1cm}\\
\begin{tabular}{|c|c|c|c|}
\hline
($0,0,0$)& $1$ & $2$ & $3$  \\
\hline
\begin{tabular}{c}
$q_{1,1}=|q_{1,1}|$, $q_{1,2}=|q_{1,2}|$,\\
 $q_{1,3}=|q_{1,3}|$ 
 \end{tabular}
 & $|q_{1,1}|$ & $|q_{1,2}|$ & $|q_{1,3}|$
\\\hline
 \begin{tabular}{c}
$q_{1,1}=|q_{1,1}|$, $q_{1,2}=|q_{1,2}|$, \\
$q_{1,3}=-|q_{1,3}|$ 
 \end{tabular}
  &$\left[\frac{{\mc I}+|q_{1,1}|-|q_{1,2}|-1}2\right]_-$ & $\left[\frac{{\mc I}+|q_{1,2}|-|q_{1,1}|}2\right]_-$
  & $0$
\\\hline
 \begin{tabular}{c}
$q_{1,1}=|q_{1,1}|$,\\
 $q_{1,2}=-|q_{1,2}|$, $q_{1,3}=-|q_{1,3}|$ 
 \end{tabular}
  &$0$ & $0$ & ${\mc I}$
\\\hline
\end{tabular}
\vspace{0.1cm}\\
\begin{tabular}{|c|c|c|c|}
\hline
($2,0,-2$)& $1$ & $2$ & $3$  \\
\hline
 \begin{tabular}{c}
$q_{1,1}=|q_{1,1}|$, $q_{1,2}=|q_{1,2}|$, \\
$q_{1,3}=-|q_{1,3}|$ 
 \end{tabular}
  &$\left[\frac{{\mc I}+|q_{1,1}|-|q_{1,2}|+1}2\right]_-$ & 
  $\left[\frac{{\mc I}+|q_{1,2}|-|q_{1,1}|-1}2\right]_+$ & $0$
\\\hline
 \begin{tabular}{c}
$q_{1,1}=|q_{1,1}|$,\\
 $q_{1,2}=-|q_{1,2}|$, $q_{1,3}=-|q_{1,3}|$ 
 \end{tabular}
 &${\mc I}$ & $0$ & $0$
\\\hline
\end{tabular}\vspace{0.5cm}\\
\begin{tabular}{|c|c|c|c|}
\hline 
($3,3,1$)& $1,1$ & $1,2$ & $2,1$  \\
\hline
 \ & $|q_{1,1}|$ & $|q_{1,2}|$ & $|q_{2,1}|$  \\\hline
\end{tabular}
\vspace{0.1cm}\\
\begin{tabular}{|c|c|c|c|}
\hline
($1,1,-1$)& $1,1$ & $1,2$ & $2,1$  \\
\hline
 \begin{tabular}{c}
$q_{1,1}=|q_{1,1}|$, $q_{1,2}=|q_{1,2}|$, \\
$q_{2,1}=-|q_{2,1}|$ 
 \end{tabular}
  &$\left[\frac{{\mc I}+|q_{1,1}|-|q_{1,2}|-1}2\right]_-$ & 
  $\left[\frac{{\mc I}+|q_{1,2}|-|q_{1,1}|-1}2\right]_-$ & $0$
\\\hline
 \begin{tabular}{c}
$q_{1,1}=|q_{1,1}|$,\\
 $q_{1,2}=-|q_{1,2}|$, $q_{2,1}=-|q_{2,1}|$ 
 \end{tabular}
 &${\mc I}$ & $0$ & $0$
\\\hline
\end{tabular}
\vspace{0.1cm}\\
\begin{tabular}{|c|c|c|c|}
\hline
($-1,-1,1$)& $1,1$ & $1,2$ & $2,1$  \\
\hline
 \begin{tabular}{c}
$q_{1,1}=|q_{1,1}|$, $q_{1,2}=|q_{1,2}|$, \\
$q_{2,1}=|q_{2,1}|$ 
 \end{tabular}
  &$|q_{1,1}|$ & 
  $|q_{1,2}|$ & $|q_{2,1}|$
\\\hline
 \begin{tabular}{c}
$q_{1,1}=|q_{1,1}|$,\\
 $q_{1,2}=-|q_{1,2}|$, $q_{2,1}=|q_{2,1}|$ 
 \end{tabular}
 &$\left[\frac{2{\mc I}-2|q_{1,1}|+|q_{2,1}|-1}3\right]_- $ 
&$0$& $\left[\frac{{\mc I}+2|q_{1,1}|-|q_{2,1}|-2}3\right]_- $ 
\\\hline
\begin{tabular}{c}
$q_{1,1}=-|q_{1,1}|$,\\
 $q_{1,2}=-|q_{1,2}|$, $q_{2,1}=|q_{2,1}|$ 
 \end{tabular}
 &$0$& 
 $0$ & ${\mc I}$
\\\hline
\end{tabular}
\end{tabular}
\caption{\footnotesize{Number of fermions localized at the various fixed points for $T^2/\Z_{2,3,4}$. When a number in the table is negative, it has to be replaced with $0$, while the other non-vanishing number on the same raw must be replaced with ${\mc I}$. The symbols $[\ldots]_{\pm}$ are used; $[x]_{-}$ is the usual integer part of $x$ corresponding, when the argument is positive, to the maximum integer which is smaller or equal than $x$; $[x]_{+}$, on the contrary, is the minimum integer which is greater or equal than $x$.}}
\end{center}
\end{table*}
In a straight--forward, but tedious way, all orbifolds with all twists could be discussed. The number of localized fermions at each fixed point, for different patterns of effective orbifold projections is summarized in Table~2 for the $T^2/\Z_{2,3,4}$ orbifolds. In the table, ${\mc I}\ge0$ is assumed, so that all states are $L$--handed, but the spectrum of $R$--handed localized states one gets when ${\mc I}\le0$ is easily obtained by inverting the signs of $p_{l,i}$ and $q_{l,i}$. For $T^2/\Z_2$ and $T^2/\Z_3$, the result for the most general pattern of effective orbifold projection $p_{l,i}$ can be obtained from the table by interchanging the fixed points. For $T^2/\Z_4$, on the contrary, the absence of discrete Wilson line, $T_{1,U}=1$ in Eq.~(\ref{tras}), is assumed. The case $T_{1,U}=-1$, and the results for the $T^2/\Z_6$ orbifold as well, which are not shown, could be easily worked out. When looking at the table, the index theorem (\ref{ind}) may sometimes appear not to be respected, one state being missing, even though the usual bulk zero--modes are correctly added to the counting. When this happens, a doubly localized state ---which is not counted in the table--- appears and makes the index theorem respected.

\end{document}